\documentclass[preprint,5p,a4paper,11pt]{elsarticle}

\usepackage{lineno,hyperref}
\modulolinenumbers[5]
\usepackage{xcolor}
\usepackage{mathptmx}      % use Times fonts if available on your 
\usepackage[linesnumbered,ruled,vlined]{algorithm2e}
\usepackage{wrapfig}

\journal{arXiv.org}

\begin{document}

\begin{frontmatter}
	
\title{Model-Based Framework for exploiting sensors of IoT devices using a Botnet: A case study with Android}

\author[1]{Zubair Khaliq}\ead{zikayem@gmail.com}
\author[1,3]{Dawood Ashraf Khan}\ead{dawood.khan@uok.edu.in}
\author[2]{Asif Iqbal Baba}\ead{ababa@tuskegee.edu}
\author[1]{Shahbaz Ali}\ead{shahbazali03@gmail.com}
\author[1]{Sheikh Umar Farooq}\ead{suf.cs@uok.edu.in}

\address[1]{University of Kashmir, Srinagar, Jammu and Kashmir, India}
\address[2]{Department of Computer Science, Tuskegee University, Tuskegee, Alabama, USA}
\address[3]{hyke.ai}

\begin{abstract}
\textit{Botnets have become a serious security threat not only to the Internet but also to the devices connected to it. Factors like the exponential growth of IoT, the COVID-19 pandemic that's sweeping the planet, and the ever-larger number of cybercriminals who now have access to or have developed increasingly more sophisticated tools are incentivizing the growth of botnets in this domain. The recent outbreak of botnets like Dark Nexus (derived from Qbot and Mirai), Mukashi, LeetHozer, and Hoaxcalls, etc. shows the alarming rate at which this threat is converging. The botnets have attributes that make them an excellent platform for
malicious activities in IoT devices. These IoT devices are used by organizations that need to both innovate and safeguard the personal and confidential data of their customers, employees, and business partners. The IoT devices have built-in sensors or actuators which can be exploited to monitor or control the physical
environment of the entities connected to them thereby violating the fundamental concept of privacy-by-design of these devices.
In this paper, we design and describe a modular botnet framework for IoT. Our framework is communication channel independent because it utilizes various available communication channels for command and control of an IoT device. The framework uses an enhanced centralized architecture associated with a novel `Domain Fluxing Technique'. The proposed framework will provide insights into how privacy in IoT devices can be incorporated at design time to check the sensors and actuators in these devices against malicious exploitation consequently preserving privacy. This paper includes design considerations, command and control structures, characteristics, capabilities, intrusion, and other related work. Furthermore, proof of concept Botnet is implemented and explained using the developed framework. }

\end{abstract}


\end{frontmatter}

%\linenumbers
\section{INTRODUCTION}
\label{intro}
%\begin{minipage}{\columnwidth}
Botnets are not new to computer systems and have been there since the 1990s. Botnets have evolved with the evolving technology and have adapted to diverse network types. The botnets were originated in PC networks; however, they have started evolving to other networks like the Internet of Things (IoT). The reason for this evolution in IoT is due to the following reasons: (1) devices have become more and more popular because of their size, versatility, and increasing computing capability, which makes them equally comparable with their PC counterparts, (2) IoT networks have become the source of information and data gathering, (3) IoT networks are designed with open source software and code is often reused to build new devices, and (4) have limited resources so they do not have malware detection or advanced security features and, when they are connected directly to the Internet, they are typically not subjected to bandwidth limitations or firewall filtering. These devices have become the point of gathering and disseminating user-related data and exploiting them for data, and privacy theft is an incentive for cybercriminals. This leads to the proliferation of botnets in these networks.
%\end{minipage}\\

\textbf{\textit{Related Work:}} 
Studies have shown how powerful attacks have disrupted the functioning of IoT devices. The type of attack determines security threats to IoT. At the top level, attacks can be functionally categorized into \textit{privacy, integrity, availability} and a second classification based on the means of launching an attack can be \textit{physical, software or side-channel}, as elaborated in \hyperref[R1]{[1]}. The above-listed categories are often used in conjunction to achieve the desired objectives. Physical attacks include - micro probing, e.g., on a circuit board, physical attacks can be launched by using probes to eavesdrop on inter-component communications \hyperref[R1]{[1]}. However, for a system-on-chip, sophisticated micro probing techniques become necessary \hyperref[R2]{[2,3]} - reverse engineering \hyperref[R4]{[4]} - eavesdropping \hyperref[R5]{[5,6]}. Software attacks are the most common form of attacks witnessed and are the major threat to IoT security.  These attacks are implemented through malicious agents such as viruses, worms, trojan horses, etc. The attackers look for vulnerabilities that provide them direct access to the system \hyperref[R7]{[7]}. A typical example is the buffer overflow problem \hyperref[R8]{[8]}. “Side-channel attacks” are attacks that are based on “Side-channel information”. Side-channel information is information that can be retrieved from the encryption device that is neither the plaintext to be encrypted nor the ciphertext resulting from the encryption process\hyperref[R9]{[9]}. Side-channel attacks include - Power Analysis Attacks both SPA(Simple Power Analysis) \hyperref[R10]{[10,11]} and DPA(Differential Power Analysis) \hyperref[R12]{[12]} which analyze the power consumption of a device and deduce information about the operations that take place and the involved parameters. Timing Attacks are a way of obtaining a user's private information by carefully measuring the time it takes the user to carry out cryptographic operations \hyperref[R13]{[13,14,15]} - Fault Injection Attacks try to induce faults in the system's hardware or software to compromise the security of the system \hyperref[R16]{[16,17]}.  Electromagnetic Analysis Attacks attempt to measure the electromagnetic radiation emitted by a device to reveal sensitive information \hyperref[R18]{[18,19,20,21,22]}. A lot of effort is being dedicated to studying the security of IoT networks, and some countermeasures have also been provided. Our work is also similar in the sense that it is a juxtaposition of these types of attacks.\\

\textbf{\textit{Problem statement: }}
In recent years we have seen real-life examples of botnets that have targeted various IoT devices like CCTV cameras, smartphones, PDAs,  routers, set-top boxes, etc. According to researchers at the security firm Imperva, a massive botnet attack earlier in 2019 utilized more than 400,000 IoT-connected devices over the course of just 13 days targeting an online streaming application\hyperref[R23]{[23]}. The research team at Spamhaus Malware Labs identified and blocked 10,263 malware botnet controllers (C\&C) hosted on 1,121 different networks in 2018, which is an 8\% increase from the number of botnet C\&Cs seen in 2017 \hyperref[R24]{[24]}. According to Symantec, a mysterious piece of software called Wifatch has been infecting tens of thousands of Linux-based home routers \hyperref[R25]{[25]}. A similar trend was found when attackers compromised more than 25,000 digital video recorders and CCTV cameras and used them to launch distributed denial-of-service (DDoS) attacks against websites\hyperref[R26]{[26]}. LizardStresser, the DDoS malware for Linux systems written by the infamous Lizard Squad attacker group, was used to create over 100 botnets, some built almost exclusively from compromised Internet-of-Things devices\hyperref[R27]{[27]}. Softpedia News is quoted as saying, "There are at least 40,000 unique IP addresses launching brute-force attacks against Telnet ports daily, and most of these IPs belong to IoT devices" \hyperref[R28]{[28]}. In 2012 an unorthodox botnet called \textit{"Carna"} compromising 420,000 small devices was used by hackers to map the Internet. From the wide range of devices at hand, we see tremendous growth in mobile botnet implementations. Mobile botnets are presently posing a serious concern to both the end-users and cellular networks \hyperref[R29]{[29]}. A general rule as a botmaster is, the bigger the botnet, the better it is. As such, attackers focus more and more on smartphone mobile phones (as about a billion-plus devices are vulnerable to being exploited).

\textbf{\textit{Challenges:}}
Implementing a botnet on IoT poses specific challenges (1)The power consumption, because if the bot consumes unusual power than expected, then it is likely to be noticed by the user (2) users are sensitive to data costs (3) some IoT devices like smartphones have the dynamic IP address allocation mechanism which may pose a serious challenge to build a botnet (4)the behaviour and structure of network traffic should be such that it should not come under any suspicion. Many studies on botnets and most of them concentrate on how a botnet can be implemented using various overlay network topologies such as centralized, distributed, P2P, TOR, etc. These papers aim to show the different mechanisms through which a botnet can sustain its C\&C(Command and Control) channel in the network systems.  These studies show how a botnet can stealthily siphon the data stolen from the victim's devices and use it for malicious purposes.

\textbf{\textit{Our Contribution: }}
This paper presents a component-based generic botnet framework that can be used to develop bot programs for any networked device. We use the proposed botnet for exploiting the sensors of an IoT device to steal private data. The network architecture we propose uses centralized network topology incorporated with the domain fluxing technique. We use a master device to publish commands on C\&C servers, which are then retrieved by the bot. The bot framework we propose can adapt to any network interface for communication purposes. Our framework can create or recreate any bot, new or old with advanced and customized features. The bot program can be customized for a specific network device, and thus we can create a botnet in any IoT network. We also discuss our botnet's level of stealthiness and resilience to attacks. The framework is designed using Model-Based System Engineering (MBSE).
% as such we abstract the overlay network topology implementation.

The main contributions of the work are as follows:
\begin{itemize}
	\item We provide a detailed architecture of the network topology of the botnet that is typically a hybrid network with multiple network interfaces at its disposal. 
	\item We propose a novel Domain Generation Technique (DGT) more efficient than state-of-art DGT.
	\item We propose a generic model-based framework for creating customized bots to exploit sensors of IoT devices. 
	\item As a proof of concept we implement a bot program from the proposed framework.	
	\item We evaluate and enumerate our implementation and provide some defense recommendations.
\end{itemize}
\textbf{\textit{Organization: }}The rest of the paper is organized as follows. Section II details the proposed architectural design: the network architecture and the bot framework. We present a case study of the proposed architecture in Section III and evaluate its effectiveness and performance. Section IV discusses the future work and conclusion.

\section{MODEL}
\label{model}
%\begin{minipage}{\columnwidth}
In this section, a detailed model of the botnet framework is presented. The botnet design requires three main components namely,\textit {Command \& Control Mechanism (C\&C)}, \textit{Payload}, and \textit{Vector}. The efficiency and reliability of a botnet are determined to a great extent by evaluating how well these components have been designed. A brief overview of these components before proceeding with our proposed model.\\ 
%\end{minipage}

\textbf{Command \& Control Mechanism:}
The Command-and-Control is the central component of a botnet. It is responsible for controlling the bots. It acts as an interface between the Botmaster (entities used to disseminate commands)  and the bots (entities for which commands are intended). Many important functions, such as load balancing, are implemented within this component. The overall efficiency, resilience, self-adaptability and stealthiness of a botnet greatly depends upon this component. \\

\textbf{Payload:}
The payload is usually a piece of malicious code that is injected into the target device/host. The injection and successful execution of a payload on the target opens a backdoor for a botmaster to access the otherwise restricted resources of the device using a bot. The payload is specific to the type of infection. For instance, if the purpose of infecting the device is data theft, the payload may collect the data from the device memory and forward it to the botmaster. Moreover, if the purpose of infection is to form a DDOS network, the malware will make the device i:e Bot (malicious code running on a device will be simply called a bot from here onwards) to send frequent data packets to the targeted infrastructure.\\ 

\textbf{Vector:}
The vector is a piece of code that enables a payload to get access to the services and physical components on an infected device. A vector is a component already installed on the device or it is a component installed on the device along with the payload. In the first case, the payload breaches the already existing component and appears as its sub-component. While as in the second case, the payload is installed along with the delivered component. In some cases, a payload may not require a separate vector, but may on its own escalate the authentication service of a device and install itself as a background service. 

In order to design the payload, we provide a general explanation of the framework of the bot. However, the design of a vector depends on factors like device operating system, purpose, and resilience of the device to attacks. Therefore, the design and implementation vary from one bot to another, making it inappropriate to propose a generalized design strategy for vectors.

To achieve the generality of the design, the Model Based System Engineering (MBSE) approach is used for the development of the framework. With MBSE we provide an integrated, coherent, and consistent system model. The system model will serve as a central repository for design decisions; each design decision will be captured as a model element in a single place within the system model. The advantage of using MBSE is to achieve the portability of the bot over various network device architectures. Once a concrete high-level model of a bot has been designed, a sufficiently robust tool is used to transform the model into an executable code, targeting a specific IoT device. This will enable our framework to produce bots with customized features with a rapid turnaround time.

At the heart of MBSE is the use of a modelling language for designing purposes, in this work we use SysML (System Modeling Language) as our modelling language to describe our framework. A SysML is a general-purpose modelling language for system engineers, its semantics are more flexible and expressive as they support various analysis schemes. Moreover, the SysML model of the bot, makes the framework to be easily understood, and the implementation can be done for any platform. The diagrams used in this model are quite intuitive and thus the framework can be understood by a novice computer security scholar easily.

\subsection{Enhanced Botnet Architecture}
\label{BBArch}
\begin{figure}[!h]
\centering
\includegraphics[scale= 0.35]{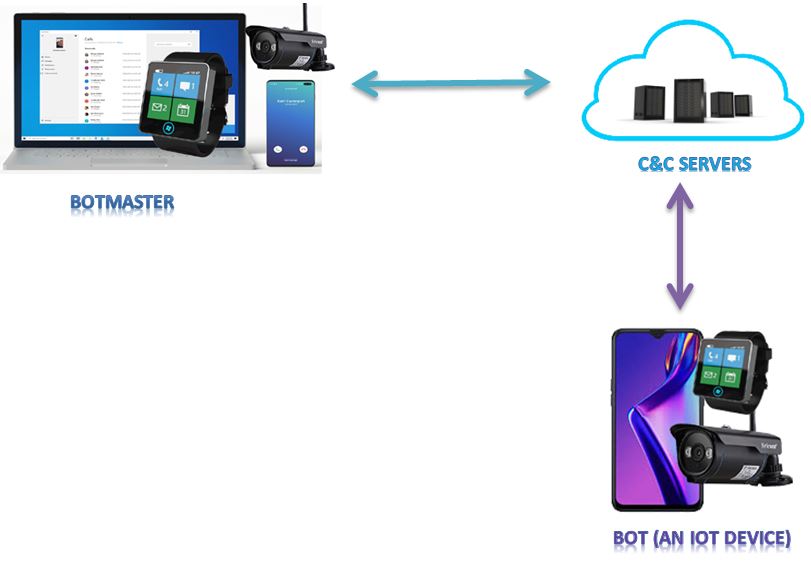}
\caption{A basic C\&C topology using multiple servers.}
\label{Genbot}
\end{figure}

We will first explain the network architecture of the botnet. A centralized network topology is used in which each device connects to a certain server to download the commands. The botmaster sits behind C\&C servers and disseminates commands to bots through the servers. The bots download their respective commands from these servers. A general overview of the network is shown in \hyperref[Genbot]{[Fig. 1]}.\\

The bots use Domain Fluxing Technique to connect with the C\&C servers to download commands and upload data. In domain fluxing an infected device, continuously changes the servers for stealthiness(to be noticed by the user) and resilience(client load balancing). This approach to a centralised system is much more efficient and easy compared to an IRC or a P2P based network.  
For this approach to work efficiently on IoT devices, we have to ascertain that this approach minimises the complexities given low-end IoT devices. In order to ascertain this, we enhance Domain Fluxing Technique (DFT) for bots to connect to newer servers. 

A Domain Generation Algorithm (DGA) generates a series of domain names over various Top Level Domains (TLD) provided seed and date as input. A similar DGA is used by both the server, and the bot with the same inputs so that they share the same domain list. Usually, about 10,000 domains are produced at one time however, the difference is in the operation of selecting the domains to be used.
 We use the following notations to represent different entities:\\
%\textit{\textbf{Assumptions:}}\\
\hspace*{5mm}$\alpha$ denote the range of domain names generated.\\
\hspace*{5mm}$\beta$ denote the number of windows.\\
\hspace*{5mm}$\gamma$ denote the number of domains in a particular window hence, denotes the number of domains the bot accesses.\\ 

\textbf{Server Operation:} Apart from sharing the DGA both server and client share the number $\beta$. A server simply bifurcates the domain range into a set of windows using $\beta$ and then tries to register a random domain in each window. The benefits of this windowing approach in terms of stealthiness and power consumption will be reaped by the client which we discuss in the next section\\

\textit{Algorithm 1: Server registers a random domain in each window using $\beta$.}
\begin{algorithm}
\DontPrintSemicolon % Some LaTeX compilers require you to use \dontprintsemicolon instead
\KwIn{List of domain names provided by DGA}
\KwOut{Registered Domain List}
$W \gets Window$\;
\For{$i \gets 0$ \textbf{to} $\beta$} {
    RegRandDomain(Wi);
}
\caption{{\sc Registers} Domains}

\end{algorithm}\\

\textbf{Client/Bot Operation:} Under normal circumstances the client has to continuously poll to each of the generated domains and tries to connect to one of the domains. The problem, however, is the time and computation required by the DFT. This is because on average a client connects to half of the domains generated. To mitigate this, we again use the windowing concept and allow the client to divide the domain range into a set of pre-specified windows $\beta$, where each window has a set of domain names $\gamma$. On the client-side, our DFT, then, selects a window randomly and polls the domains inside the window until it finds one that has been registered by the server. If a client does not find a C\&C server in a window it again selects a random window and polls its domains. It keeps doing that until it finds a C\&C server meant for it and after receiving an acknowledgement from the C\&C server it saves its address for further communication. We call this Enhanced Domain Fluxing. \\

\textit{Algorithm 2: Returns a single domain from a list of domain names using Enhanced Look-Up assuming the server to have registered a domain in each window.}

\begin{algorithm}
\DontPrintSemicolon % Some LaTeX compilers require you to use \dontprintsemicolon instead
\KwIn{List of domain names provided by DGA}
\KwOut{A single domain}
    $RandWindow \gets SelRandWindow(\beta)$\;
    $\gamma \gets \alpha \div \beta$\;
    \For{$j \gets 0$ \textbf{to} $\gamma$} {
        $ack \gets PollServer(i,j)$\;
        \If{$ack = positive$} {
            \Return{$PollServer(i,j)$}\;
        }
    }
\Return{RegDomainFailed}\;
\caption{{\sc Returns} a Single Domain}
\end{algorithm}

Using an example we show that the windowing concept discussed above gives the client considerable leverage in terms of stealthiness and power consumption. This is required for the client because it is a low-end IoT device. We consider stealthiness by measuring the amount of bandwidth consumed because if our bot consumes low bandwidth then it is almost impossible to be noticed by the user. To determine power consumption we consider the time required by the bot to identify a domain. 

\textbf{Example 1:} Let\\
\hspace*{5mm}If $\alpha$ = 10,000.\\
\hspace*{5mm}$\beta$ = 100.\\
    
\textit{Bandwidth and Time Consumption:}Without using the windowing concept a bot has to exhaustively search the domain list to identify a server. Assuming an average case for this linear search, a bot requires about $\gamma$ = $\alpha$ $\div$ 2 = 10,000 $\div$ 2 = 5,000 domains to be accessed. Assuming each domain access(an HTTP request) to consume a bandwidth of about 1500bytes, and time of about 200ms,  this approach consumes approximately 5000 * 500Bytes which is about \textit{2442KBs} of bandwidth, and 5000 * 200ms =  \textit{1000s} of time .\\
Now using the windowing concept just discussed, assuming the server to have registered a domain in each window we require each bot to access $\gamma$ = $\alpha$ $\div$ $\beta$ = 10,000 $\div$ 100 = 100 domains to be accessed. Applying same assumptions this approach consumes 100 * 500Bytes which is about \textit{49KBs} of bandwidth, and 100 * 200ms = \textit{20s} of time. From the difference in derived numbers of both approaches, it is clear that the proposed windowing approach has a considerable advantage over its former counterpart.

\begin{figure}[!h]
\centering
\includegraphics[scale= 0.25]{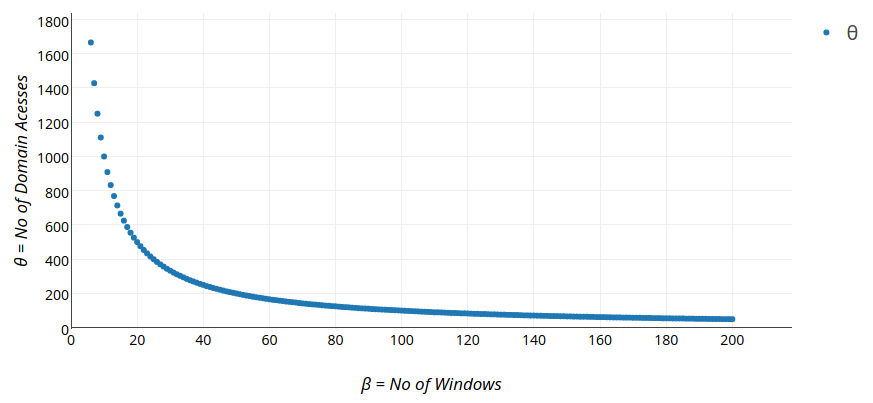}
\caption{Graph showing accessed domains by a bot with $\alpha$ = 10,000.}
\label{DGA}
\end{figure}

One point that is worth mentioning here is that the server has to register $\beta$ domains, so a balance is to be maintained because of $\beta$ and $\gamma$ being inversely proportional to each other as seen from the graph plotted below\hyperref[DGA]{[Fig. 2]}.

The more the number of windows the lesser no of domains registered resulting in a much narrow range of domains to be accessed by the bot. All of these values are to be adjusted for optimality as per the resources available.\\ 

Now we show the steps of how a bot joins the network and later downloads the commands from the C\&C server.\\

\begin{figure}[!h]
\centering
\includegraphics[scale =0.35]{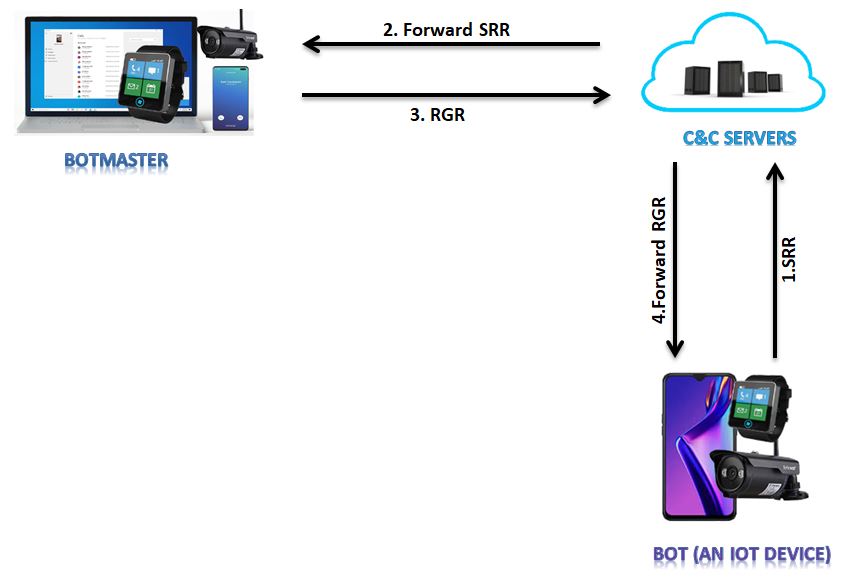}
\caption{Registration of a bot.}
\label{BotReg}
\end{figure}

\textbf{Registration:} For the first time, the bot sends a \textit{Special Registration Request} (SRR) associated with details that identify the device uniquely to the C\&C server it is currently connected with, which in turn forwards the same to the Botmaster. The Botmaster then stores the information in its database and acknowledges the C\&C server with \textit{Registration Granted Response} (RGR). The C\&C then saves the bot details in its database and forwards the response to the bot. In this way the bot only knows the details of the C\&C server without knowing the details of the botmaster, hence providing added security to the botmaster. The bot now is said to be registered on the botnet\hyperref[BotReg]{[Fig. 3]}.

\textbf{Command Dissemination and Data Upload:} Since only the botmaster has complete detail of the botnet, it disseminates commands for the bots via the C\&C servers. Firstly The botmaster publishes the commands on the C\&C servers. The C\&C servers then save the commands and wait until the bot requests the same. The bot then sends a request called \textit{Download Command Request} (DCR) to the C\&C server and if there is a command available for this bot it allows the bot to download the same. Since the IP address allocation system on many IoT devices is dynamic therefore we uniquely identify a device by its device id. This helps us to relocate the IP address of a device if it has changed via RCIPB(Request to change IP of Bot). The botmaster provides the C\&C server with a \textit{List} which contains the id and IP of the devices to which commands are to be sent. The C\&C server checks each DCR request against its corresponding entry in the \textit{List}. It allows the bot to download a command if there is one else it sends the bot a message "Nothing for You".\\
\textit{Algorithm 3: C\&C server sends a command to the bots.}
\begin{algorithm}
\DontPrintSemicolon % Some LaTeX compilers require you to use \dontprintsemicolon instead
\KwIn{List of bots from Botmaster, DCR(BotID,BotIP) from Bot}
\KwOut{Sends a command to the bots.}
$List[][] \gets List\hspace*{1mm}of\hspace*{1mm}bots\hspace*{1mm}to\hspace*{1mm}which\hspace*{1mm}commands\hspace*{1mm}are\hspace*{1mm}to\hspace*{1mm}be\hspace*{1mm}sent$\;
$BotID \gets Id\hspace*{1mm}of\hspace*{1mm}Bot\hspace*{1mm}which\hspace*{1mm}requested\hspace*{1mm}the\hspace*{1mm}command$\;
$BotIP \gets \hspace*{1mm}Ip\hspace*{1mm} adress\hspace*{1mm} of\hspace*{1mm} Bot\hspace*{1mm} which\hspace*{1mm} requested\hspace*{1mm} the\hspace*{1mm} command$\;
\For{$i \gets 0$ \textbf{to} $length(List)$} 
    {
    $j \gets 0$\;
        \If{$BotID = List[i][j]$ \& $BotIP != List[i][j+1]$}        
        {
            $List[i] = BotIP$\;                
            $SendMsg(RCIPB,Botmaster)$\;
            $SendMsg(Command,Bot)$\;
            $Exit()$\;
        }
        \ElseIf{$BotID = List[i][j]$ \& $BotIP = List[i][j+1]$}
        {
            $SendMsg(Command,Bot)$\;
            $Exit()$\;
        }    
    }                                
$SendMsg(“Nothing for You”,Bot)$\;                                        
\caption{{\sc COMMAND} Dissemination by C\&C.}
\end{algorithm}

\begin{figure}[!h]
\centering
\includegraphics[scale=0.35]{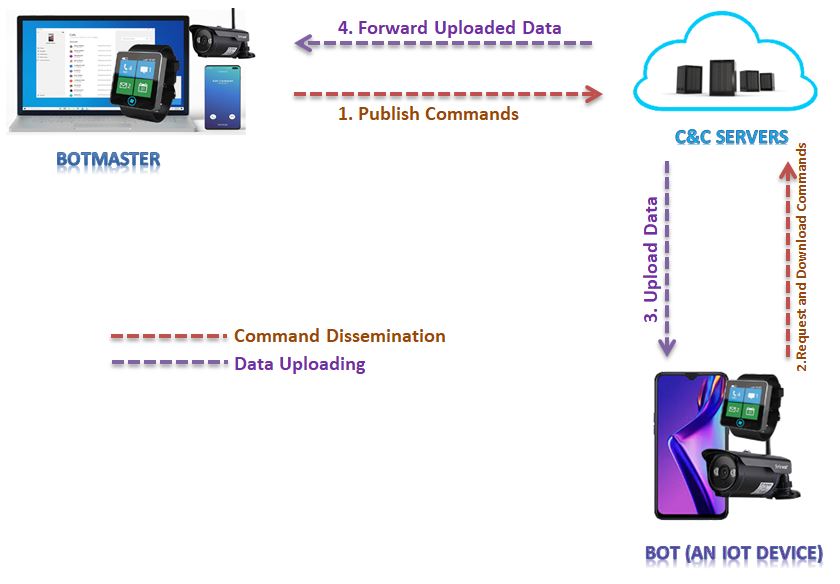}
\caption{Command Dissemination \& Data Upload}
\label{C&D}
\end{figure}

The bot then executes the command and returns the result to the server which in turn forwards the result to the botmaster and after receiving an acknowledgement for this the bot deletes the corresponding data. Note the bot pulls the commands and this pull mechanism is important because it provides for that added security as there is no activeness from the server(connection is not persistent)\hyperref[C&D]{[Fig. 4]}.\\

\hspace*{10mm}Now we try to explain the Reliability and Stealthiness of the proposed framework. Reliability here determines the availability of the network to the bots and also how the network returns to its original shape or position in case of an attack. Stealthiness determines how the botnet will run without being detected. \\
\textbf{Reliability:} We consider this issue under three main scenarios.\\
\hspace*{2mm}\textit{1. Update to new C\&C’s.}\\
\hspace*{2mm}\textit{2. Install a new botmaster in case of an attack.}\\
\hspace*{2mm}\textit{3. Register a new C\&C server in case it goes down.}\\

\textbf{\textit{1.\hspace*{2mm}Update to new C\&C’s:}}
In the domain fluxing technique every bot break-up with the current C\&C server and connects to a different C\&C server after a predetermined time. After that, the bot runs the DGA and connects to a different server. But what if the C\&C server is brought down at some point while the bot is supposed to download commands or upload data. This will not have an effect in our case because the bot connects to a different server after a pre-determined time.\\

%Framework figure pre-plotted for adjustment

\textbf{\textit{2.\hspace*{2mm}Install a new botmaster in the case of an attack.}}
We expect that at any point in time our botmaster may be compromised. However, we let the botmaster continuously replicate the information of the whole network to some offline device using a persistent segment tree which later allows the botmaster to debug the changes to the network and restore to a previous known good configuration. \\
\label{framework}
%\begin{minipage}{\columnwidth}

\textbf{\textit{3.\hspace*{2mm}Register a New C\&C Server in case it goes down.}}
We assume that our C\&C servers can be brought down. In that case, we install a new C\&C server to keep the whole botnet alive. We assume the information related to the C\&C servers is possessed by the botmaster so the botmaster will simply install a new C\&C server and update it with the state of the previous C\&C server. The new C\&C server then broadcasts a message Request to Change Address of C\&C (RCAd) to the bots. The bots will extract the new address from the RCAd message and update their respective cache with this new C\&C server.\\

\textbf{Stealthiness in Control:} Many people have proposed techniques to determine the control or behaviour of the botnets to mitigate their adverse effects using detection techniques\hyperref[R30]{[30-32]}. Major advancements for our botnet in this field come from the fact of using the enhanced version of domain fluxing. A simple technique to limit access to C\&C infrastructure is to block access to IP addresses and domains which are known to be used by C\&C servers. The disadvantage of using this technique is that it requires malware researchers to maintain an up to date list of all domains and IP addresses associated with malware. Since we change to a new C\&C server continuously so it is difficult for our botnet to come under this radar. Many DNS based detection techniques are already there which analyse the domains requested by the malware e:g Villamarín-Salomón and Brustoloni \hyperref[R30]{[30]} have shown that a host which is repeatedly requesting a domain name that doesn’t exist is more likely to contain malware. A big consequence of these techniques is that they cannot detect C\&C channels if the domain looks normal. Thus, we can avoid further detection by using DNS infrastructures similar to a regular web hoster in combination with a legitimate-sounding name for example by appending valid Dictionary names to the generated Domain Names.\\
A bot regularly sends traffic to the C\&C server to receive new commands. Such traffic is sent automatically and is usually sent on a regular schedule. The behaviour of user-generated traffic is much less regular, thus bots may be detected by measuring this regularity. AsSadhan et al. \hyperref[R31]{[31]} have proposed a system to detect hosts generating regular traffic. We overcome this technique by allowing a bot to send data at random intervals of time so that there is no regularity in the traffic. Giroire et al. \hyperref[R32]{[32]} have designed a similar detection system that measures the temporal persistence of traffic. The system attempts to find hosts which keep connecting to the same server. Bots are likely to keep connecting to the same C\&C server as long as it is online, thus persistent connections may be used to detect C\&C channels. Our framework also overcomes this as in our framework a connection with the servers is not persistent as we do server hopping.

\subsection{Modular Architecture of Bot Framework}
 \begin{figure*}[!]
 	\begin{center}
 		\includegraphics[scale = 0.25]{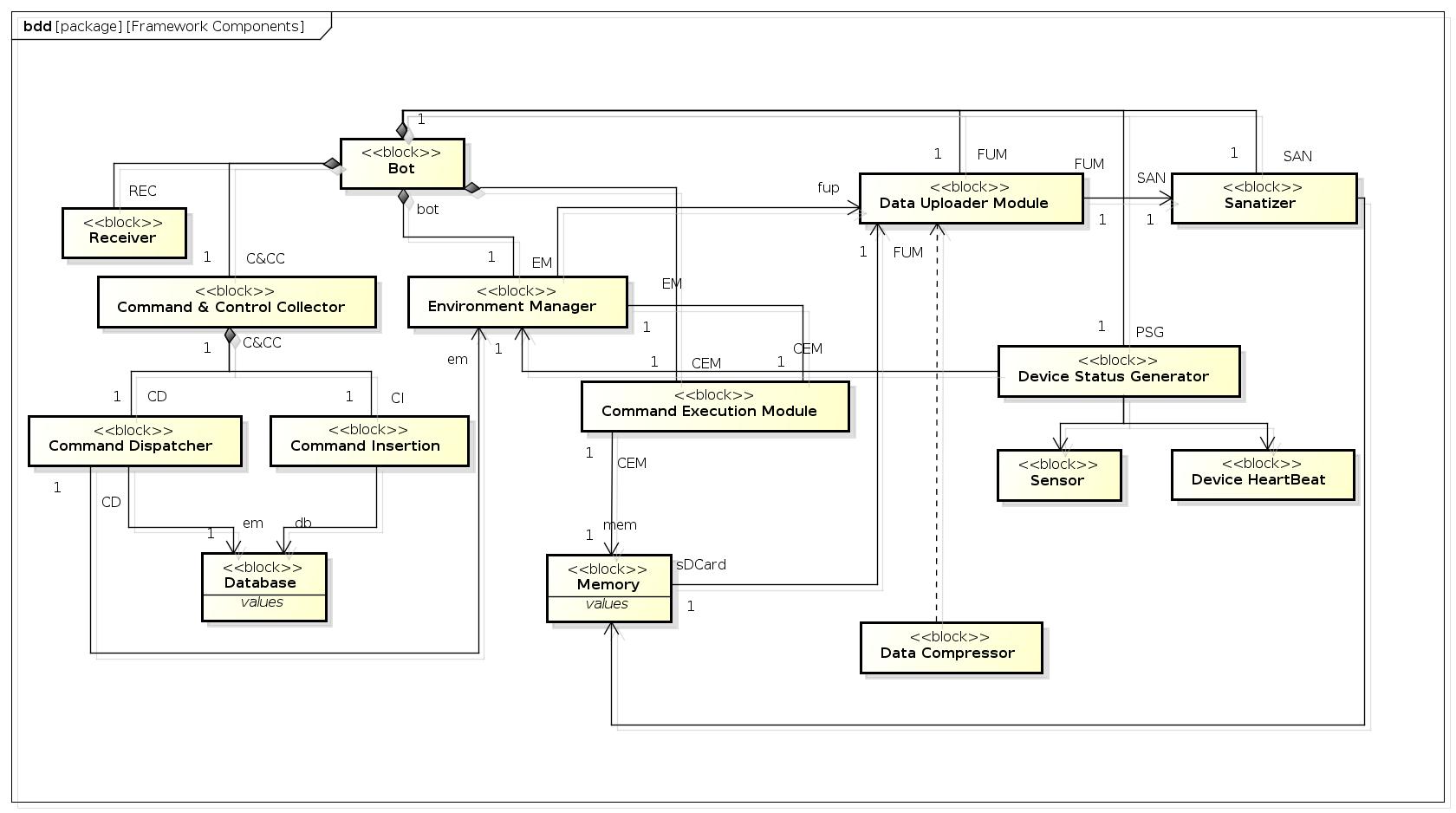}
 		\caption{Design of BOT Framework.}
 		\label{fig2}
 	\end{center}      
 \end{figure*} 

The \hyperref[fig2]{[Fig. 5]} shows the Block Definition Diagram (BDD) of the framework to highlight the design strategy. It provides a complementary view of the framework in a structured manner. The different elements or modules that constitute the BDD are called elements of the definition. The elements of definition may or may not consist of sub-modules, depending upon their architecture. The structural relationship among various modules and sub-modules, as shown in the figure \hyperref[fig2]{[Fig. 5]}, are very important as these relationships convey the levels of system decomposition and type classification. The framework suggests that the bot is internally composed of:
\begin{itemize}
\item Receiver (REC) Module
\item Command and Control Collector (CCC) Module
\item Environment Manager Module (EM)
\item Sanitizer Module (SAN)
\item Command Execution Module (CEM)
\item Phone Status Generator Module (PSG)
\item Device Heartbeat Module (DHB)
\item Sensor Module (SM)
\item File Uploader Module (FUP) and 
\item Data Compressor Module (DC).
\end{itemize}

%\end{minipage}\\

\begin{figure*}[!h]
	\begin{center}
		\includegraphics[scale = 0.35]{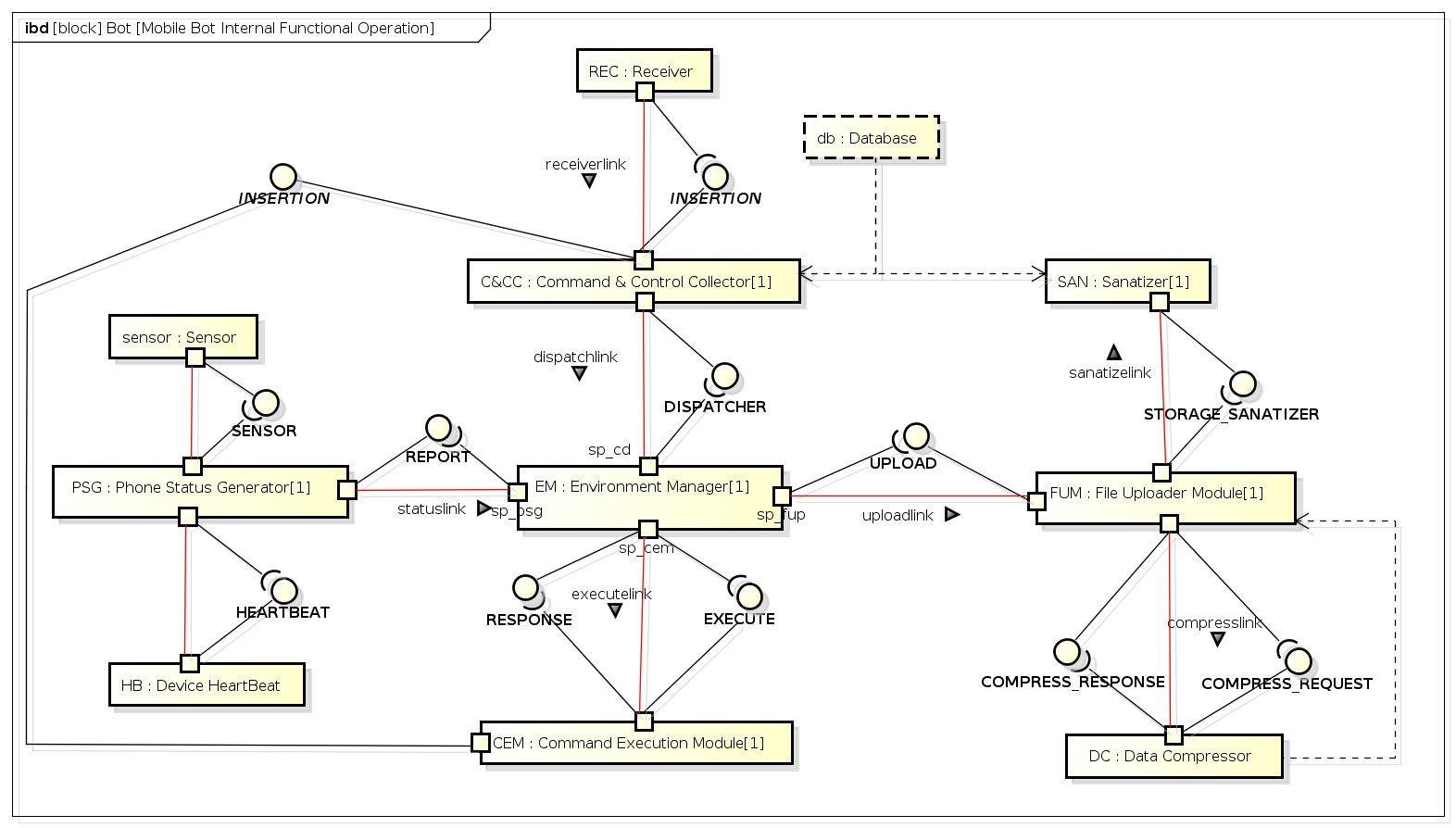}
		\caption{Architecture of the BOT Framework.}
		\label{fig3}
	\end{center}       % Give a unique label
\end{figure*}

The architecture of the BOT framework, shown in \hyperref[fig3]{[Fig. \ref{fig3}}, is based on an asynchronous client-server model, wherein the clients (bots) communicate with the server (Command-and-Control, botmaster) for receiving commands and exchanging information. The infected device can receive the commands and other information from the server through any receiving interface. Different types of receiving interfaces are SMS, Bluetooth, Wi-Fi, Infrared, and other networking interfaces. The command is received through an interface by the bot and then processed i:e have a resultant effect of the command processed. The resultant information is sent back to the server through a designated interface. The detailed design and architecture of various components and sub-components of the framework is presented in the following sub-sections.
\subsubsection{Bot Module (BM)}
\label{bm}

The \texttt{Bot module} is the root module of the framework. It is a composite block that represents a complete working bot entity, see \hyperref[fig3]{[Fig. 6]} displays seven composite associations from the \texttt{Bot} block to subsystem blocks. This association conveys that a correctly developed and assembled \texttt{Bot} will be composed of one Receiver, one Command and Control Collector, one Command Execution, one Environment Manager, one Sanitizer, one Data Uploader and one Device Status Generator module. This component represents a high-level abstraction of the payload environment and its deployment on a device categorises the device as infected. At runtime, the \texttt{Botnet} may contain several instances of a \texttt{Bot}. This means that there may be several deployment variants corresponding to an implementation of \texttt{Botnet}. Thus the objective of a \texttt{Bot} module is to distinguish every infected device on the network so that valid scalability may be achieved.

\subsubsection{Receiver (REC) Module}
\label{rec}

\begin{figure}[!h]
\centering
\includegraphics[scale= 0.30]{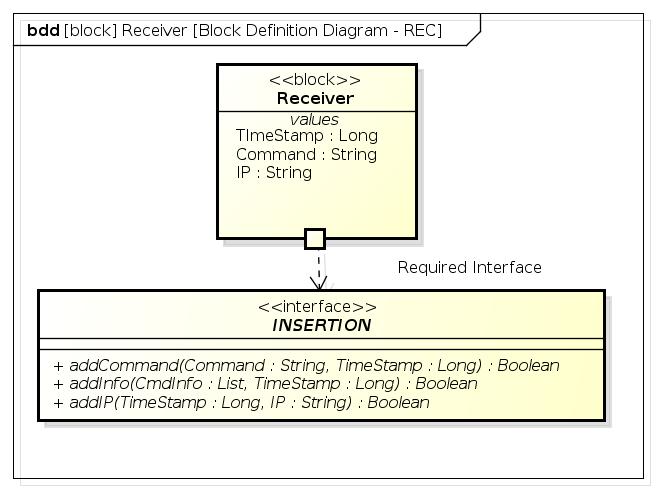}
\caption{Block Definition Diagram of Receiver (REC) Module.}
\label{fig4}
\end{figure}

The \texttt{Receiver module (REC)} \hyperref[fig4]{[Fig. 7]} is a generalized module for various network interfaces. REC is an important component of the underlying framework, its importance can be sensed from its ability in separating the communication aspect of Botnet from its functional aspect.

%\begin{figure}[!h]
%\centering
%\includegraphics[scale=0.35]{figs/BDD-REC.jpg}
%\caption{Block Definition Diagram of Receiver (REC) Module.}
%\label{fig4}
%\end{figure} 
A \texttt{Botnet} requires a communication channel to issue commands and to receive the retrieved information. In a particular device, we can have various channels of communication that can be used for command and control, and information exchange. The commonly used channels in the case of IoT devices are, Bluetooth, RFID's, Zigbee, Ethernet, WIFI, USB etc. \\
A \texttt{Bot} can use any of the above-mentioned communication channels for a connection to a remote device. While selecting a communication channel, thorough consideration must be given to the availability of the channel and the protocol overhead it produces. The capability of the framework to adapt to any communication channel is provided by the \texttt{Receiver (REC} Module, which acts as a supertype module and is inherited by the sub-modules. The sub-modules, in turn, are specialised elements like SMS, internet, Bluetooth, Wi-Fi and Voice calls. \texttt{Receiver (REC)} Module provides an abstraction layer for the framework and encapsulates the various channels available and their underlying communication mechanisms. To add a new communication channel to the framework requires defining that channel as a separate sub-module inheriting the \texttt{Receiver (REC)} Module and then sending the update to the system so that the newly added channel is available for communication and control purposes. 

\subsubsection{Command and Control Collector (CCC) Module}
\label{ccc}
\begin{figure}[!h]
\centering
\includegraphics[scale= 0.30]{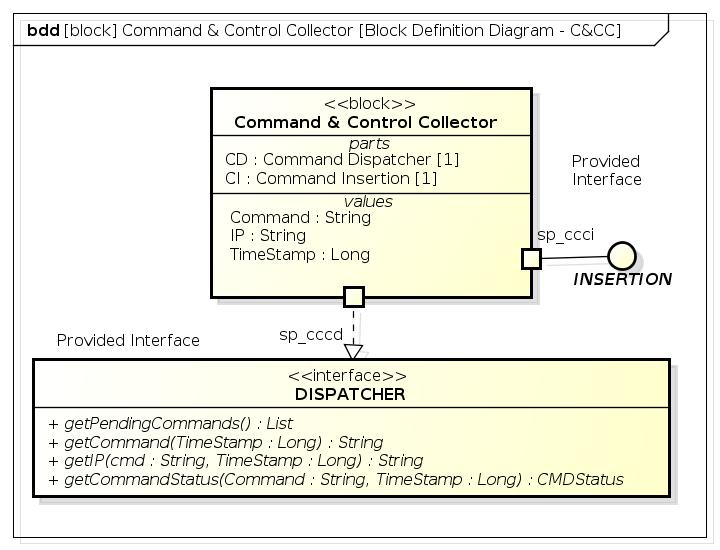}
\caption{Block Definition Diagram  of Command \& Control Collector (CCC).}
\label{fig6}
\end{figure}

The \texttt{Command \& Control Collector (CCC)}, as shown in figure \hyperref[fig6]{[Fig. 8]}, is a composite module. This module deals with the C\&C information received by the \texttt{Bot} from the botnet. When any C\&C information is received by the bot it needs to be preserved until it is processed by the \texttt{Bot} system. The objective of the \texttt{CCC} is to provide the C\&C information to the bot system as and when the demand arises, its structure and functionality can be decomposed into two separate sub-modules: \texttt{Command Insertion} and \texttt{Command Dispatcher}. A \texttt{CCC} is composed of one \texttt{Command Insertion} and one \texttt{Command Dispatcher} module. A\texttt{CCC} is among the first components that a bot needs for its proper functioning. A \texttt{CCC} is responsible for the creation and management of the database used for the storage of C\&C information in the bot.  A \texttt{CCC} also provides for the storage and acquisition of the information needed by the bot. Moreover, a \texttt{CCC} also provides mechanisms for upgrading, altering or destroying the database.

\begin{figure}[!h]
\label{figCD}
\centering
\includegraphics[width=0.50\textwidth, height= 8cm]{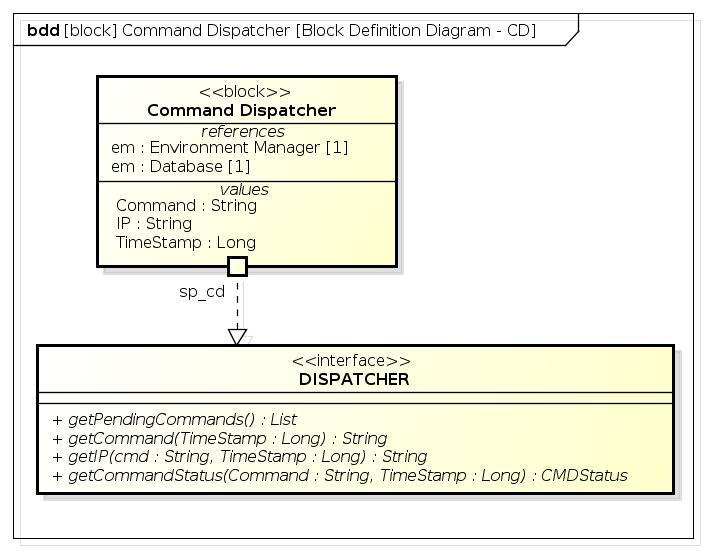}
\caption{Block Definition Diagram  of Command Dispatcher (CD).}
\end{figure}

\texttt{Command-and-Control Collector (CCC)} module has two interfaces associated with it. These interfaces are the \texttt{INSERTION} interface and the \texttt{DISPATCHER} interface. The \texttt{INSERTION} interface is used to insert data into the bot database. Similarly,  the \texttt{DISPATCHER} interface is used to retrieve the data stored in the bot database. Since the \texttt{CCC} has to provide both of these interfaces for bot utility, it delegates the duty to two separate sub-modules. These two modules are \texttt{Command Insertion (CI)} and \texttt{Command Dispatcher (CD)}.
\label{cd}
\subparagraph{Command Dispatcher Module (CD):}Command Dispatcher module \hyperref[figCD]{[Fig. 9]} deals with retrieval of data from the database. The \texttt{Command Dispatcher (CD)} provides the necessary database operations for retrieving data. Whenever there is a need for retrieving the data from the database, an instance of \texttt{Command Dispatcher (CD)} is created to handle this retrieval. \texttt{CD} module provides the \texttt{DISPATCHER} interface. This interface provides operations like \textit{getPendingCommands()}, \textit{getCommand(timestamp: Long)},  \textit{getIP()}, and \textit{getCommandStatus()}.
\label{ci}
\subparagraph{Command Insertion Module (CI):}Command Insertion module deals with inserting data into the database.
\begin{figure}[!h]
\centering
\includegraphics[width=0.50\textwidth, height= 7cm]{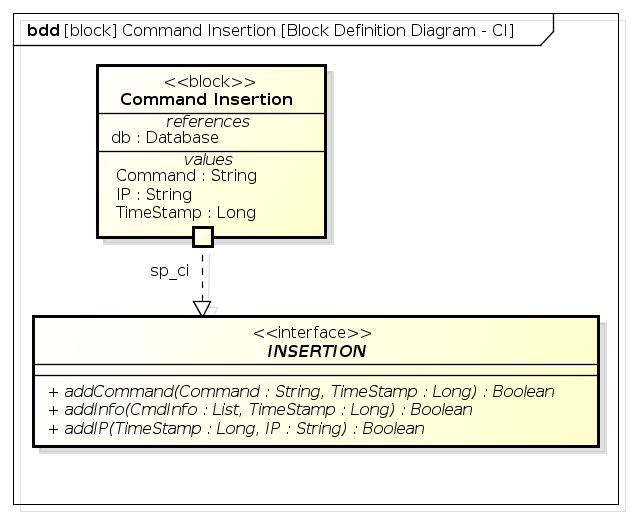}
\caption{Block Definition Diagram  of Command Insertion (CI).}
\label{figCI}
\end{figure}
\label{dm}
At any instance of time, if there is a need to add some data to the database an instance of \texttt{Command Insertion (CI)} is created to handle the insertion operation \hyperref[figCI]{[Fig. 10]}. This module provides the \texttt{INSERTION} interface. This interface provides the operations like \textit{addCommand()} and \textit{addIP()} .

\subparagraph{Database Module (DM):}A bot requires a database to store information and commands. Usually \textit{MySQLite} is used to implement this database. If MySQLite is not available on the concerned device a simple file storage operations can be utilised.

\subsubsection{Environment Manager Module (EM)}
\label{em}
\begin{figure}[!h]
\centering
\includegraphics[width=0.50\textwidth, height= 9cm]{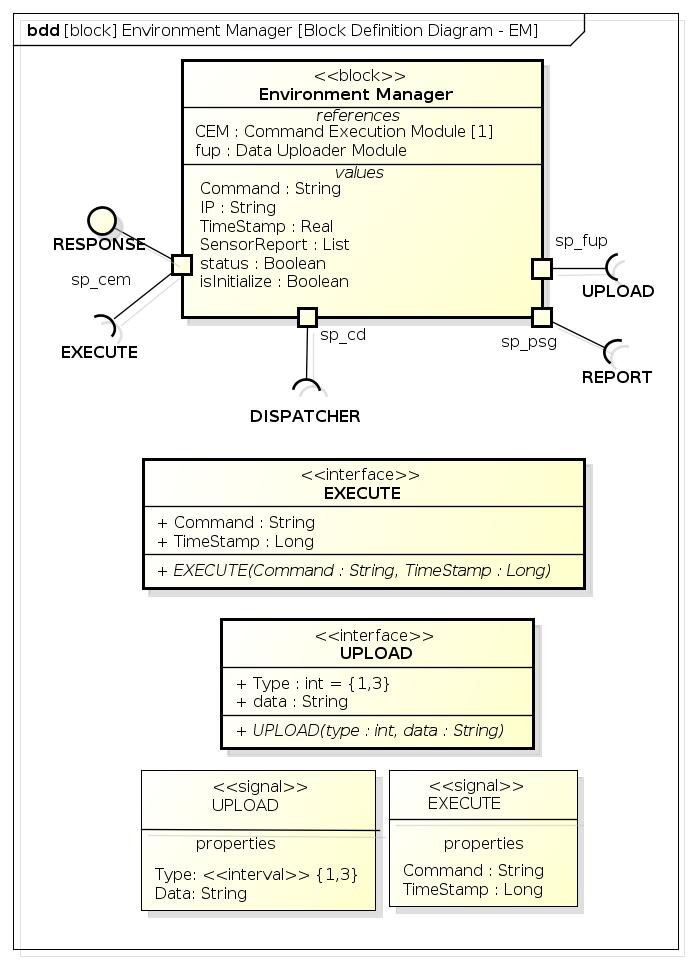}
\caption{Block Definition Diagram of Environment Manager Module (EM).}
\label{fig7}
\end{figure}

\texttt{Environment Manager Module (EM)} is the backbone of our framework. It is a versatile and robust component, which performs a lot of functions and makes many decisions regarding the execution of various processes associated with the bot. Figure \hyperref[fig7]{[Fig. 11]} gives the Block Definition Diagram of the EM. 

The importance of the EM within the model can be determined by the fact that in order to instantiate the bot, EM must be instantiated first. EM is responsible for invoking many other modules and their related operations that are necessary for the bot. EM has four ports through which it interacts with the other modules of the framework and vice-versa. Each port has an interface associated with it, which is either a required interface or provided interface. There are two reference modules for EM. The first reference module is the Command Execution Module (CEM) and the second is the Data Uploader Module (DUP). \\
EM communicates with CEM \& DUP through the two standard ports named \textit{sp\textunderscore cem} \&  \textit{sp\textunderscore fup}, the other two ports named \textit{sp\textunderscore cd} \& \textit{sp\textunderscore dsg} are meant to interact with the Command Dispatcher (CD) \& Device Status Generator (DSG) modules. The coupling of the EM with various components of the bot shows diverse functionality of the EM.\\
EM coordinates with DSG through the REPORT interface, so that it is well aware of the status of the various operations on the device which include both hardware and software operations. The DISPATCHER interface provided by CD is used for getting the C\&C information from the database. EM coordinates with CEM regarding the execution of various commands through the \textit{sp\textunderscore cem} using the EXECUTE \& RESPONSE interfaces. EM also has the responsibility to ensure that the data is being uploaded to the C\&C server by DUP correctly, for this UPLOAD interface is provided by the DUP.

\subsubsection{Device Status Generator Module (DSG)}
\label{dsg}
\begin{figure}[!h]
\centering
\includegraphics[width=0.50\textwidth, height= 8cm]{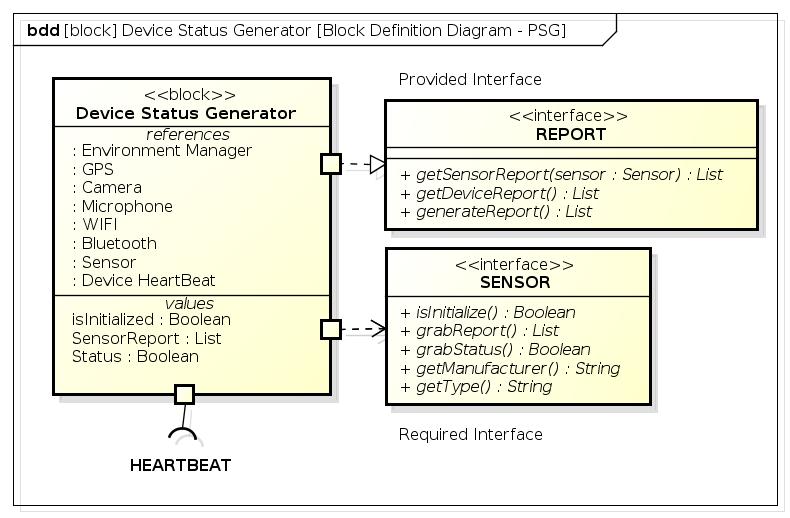}
\caption{Block Definition Diagram for Device Status Generator (DSG).}
\label{fig11}
\end{figure}

\texttt{Device Status Generator (DSG)}, shown in figure \hyperref[fig11]{[Fig. 12]}, provides the Environment Manager with information about the resources of the device. A device may be comprised of many hardware peripherals for its functioning. These peripherals may be either integrated or externally connected to the device. Information about the capabilities and features of these resources available to the device is necessary for the Bot, to exploit them for multipurpose functioning. 

The DSG in the framework is meant for this kind of work, it obtains the details regarding these resources from the onboard devices in a system and then periodically check the status of these resources. Awareness about the status and capabilities of resources of a device is important. This is because the bot is supposed to run on devices having constraints on resources and also for ensuring the availability of resources when a command is submitted for execution. This information is also needed because different devices have resources like cameras, microphones, and accelerometers manufactured by different manufacturers. Thus, they may differ in the mode of operation, and the capability of performing a certain operation. Moreover, the devices may have a heterogeneous set of available resources.\\

DSG has three standard ports associated with it among these port \textit{sp\textunderscore em} is used to report the status of various resources of the device to the EM using the REPORT interface. The remaining ports are used to interact with two reference modules \texttt{Sensor} module \& \texttt{Device HeartBeat} module. Interaction with \texttt{Sensor} module is obtained through \texttt{SENSOR} Interface which is provided by \texttt{Sensor} Module, and \texttt{HEARTBEAT} interface is used to communicate with the \texttt{Device HeartBeat} module. The DSG has a reference association with the \texttt{Sensor} module \& \texttt{Device HeartBeat} module. This means in an operational system DSG extensively communicates with Sensor \& Device HeartBeat Modules for the status generation. Since dealing with sensors, peripherals and other onboard resources is a low level(system level) job thus DSG serves as a layer of abstraction between the EM, SM, and DHB Modules.

\subsubsection{Device Heartbeat Module (DHB)}
\label{dhb}
\begin{figure}[!h]
\centering
\includegraphics[width=0.50\textwidth, height= 8cm]{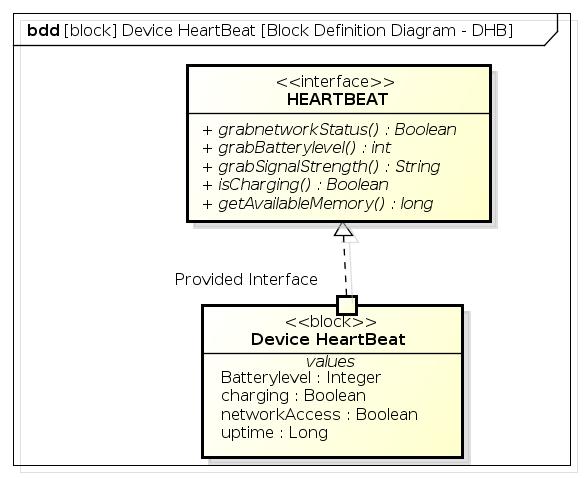}
\caption{Block Definition Diagram for Device HeartBeat Module (DHB).}
\label{fighb}
\end{figure}

\texttt{Device HeartBeat Module} \hyperref[fighb]{[Fig. 13]} is a concrete component of the framework which is delegated with a job to extract information about various onboard resources and operations that are running on the device. Resources like input power, signal strength, the number of processes running and the memory available etc. are analysed by the DHB. 

DHB is associated with one port that provides the HEARTBEAT interface, this interface is used by the DSG to get the information generated by the DHB. The DHB module may either use various device API's available for getting the resource information or a concrete implementation of the procedures can be made in this component for specific hardware if the system API is unavailable

\subsubsection{Sensor Module (SM)}
\label{sm}
\begin{figure}[!]
\label{figSM}
\centering
\includegraphics[width=.50\textwidth, height= 7cm]{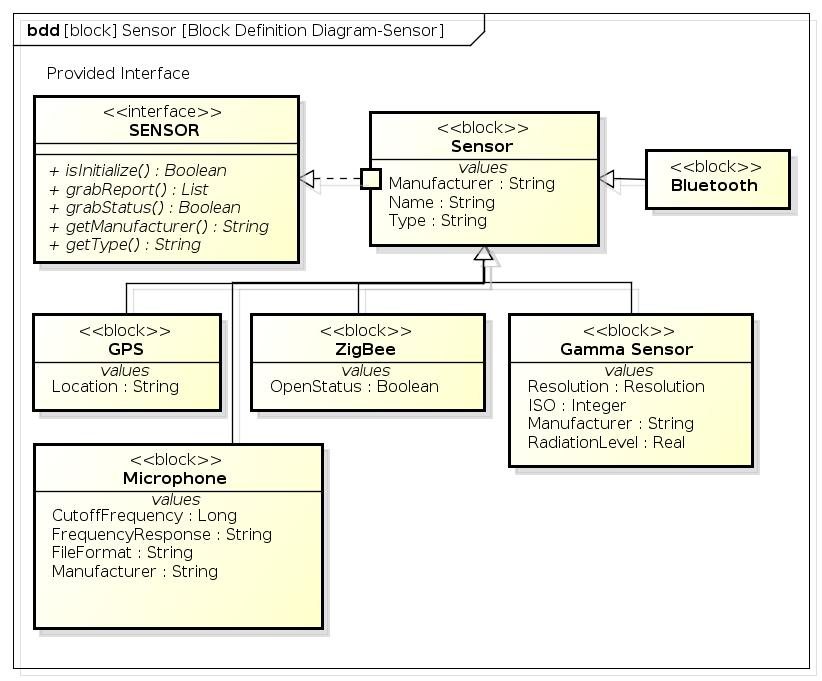}
\caption{Sensor Module (SM) Methods and Attributes.}
\end{figure}

The \texttt{Sensor Module} \hyperref[figSM]{[Fig. 14]} is another concrete component of the framework which is delegated with a job to extract status information about various device sensors, peripherals and hardware attached that are to be used by the Bot system for data gathering.

An SM is associated with a single port that provides a SENSOR interface. SM adds a layer of abstraction for various peripheral elements, it acts as a supertype and defines a common API within the model. This common API propagates down the hierarchy to all the subtypes that inherit the SM. To get details about a particular sensor (say a Camera) associated with the device, an implementation module named Camera is associated in the model. The camera module inherits the SM and implements the various operations that are provided by the SM. In this way, polymorphism is achieved for various device sensors used by the Bot system.

\subsubsection{Command Execution Module (CEM)}
\label{cem}
\begin{figure}[!h]
\centering
\includegraphics[width=.50\textwidth, height= 9cm]{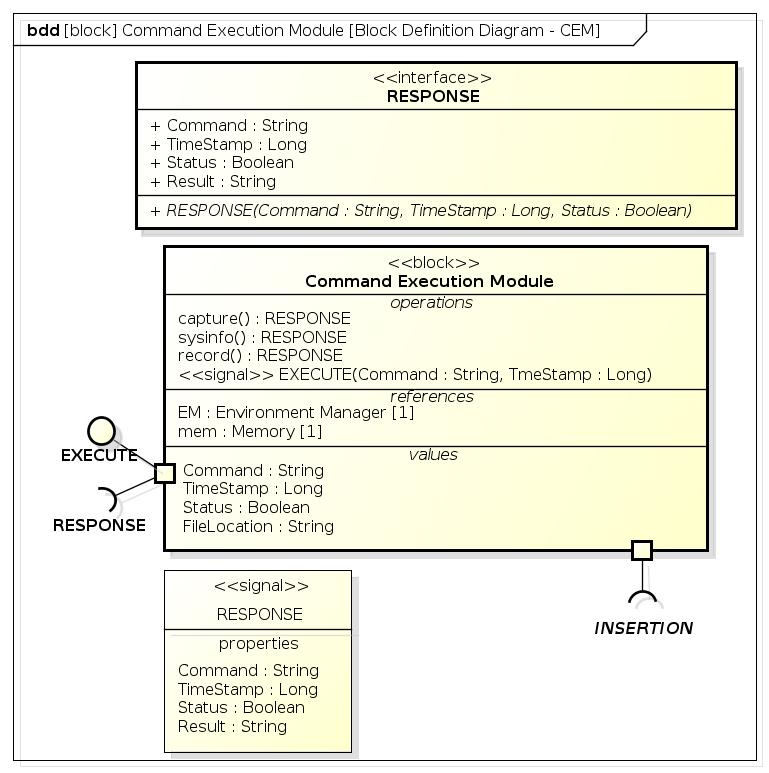}
\caption{Block Definition Diagram of Command Execution Module (CEM).}
\label{figcem}
\end{figure}

The \texttt{Command Execution Module (CEM)} \hyperref[figcem]{[Fig. 15]}  is responsible for the execution of commands on the device. The \texttt{CEM} has two standard ports associated with it \textit{sp\textunderscore em} and \textit{sp\textunderscore ci} through which it has interaction with \texttt{Environment Manager} and \texttt{Command Insertion} respectively.The port \textit{sp\textunderscore em} provides \texttt{EXECUTE} interface and requires \texttt{RESPONSE} interface for communicating with \texttt{Environment Manager}. The EM using operations of EXECUTE interface sends execute signal along with the command to the CEM, the CEM gets invoked by the execute signal and executes the corresponding procedure meant for the command, on successful completion of the command CEM sends the response signal to the EM using the RESPONSE interface. After the execution of the command, CEM uses the INSERTION interface of the Command Insertion Module for updating the STATUS of the C\&C command in the database.
\subsubsection{Data Compressor Module (DC)}
\label{dc}
The \texttt{Data Compressor module} is a dependent module. It is responsible for the compression of the data that is to be uploaded by the Bot system. It depends on the DUP module for the input data. DUP decides whether the data to be uploaded needs the service of a DC or not. The DC has two interfaces COMPRESS\textunderscore REQUEST and  COMPRESS\textunderscore RESPONSE. When DUP needs to compress the data, it sends a COMPRESS\textunderscore REQUEST signal, containing the path of the file, to the DC. The DC runs different pre-specified compression algorithms for the different types of files and then sends the new path and the name of the compressed file to the DUP, which then uploads it to the specified server. The DC module uses the COMPRESS\textunderscore RESPONSE interface to communicate back with the DUP module.  

\subsubsection{Data Uploader Module (DUP)}
\label{fum}

\begin{figure}[!h]
\centering
\includegraphics[scale =0.70]{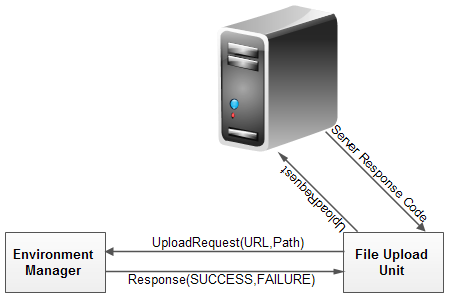}
\caption{Data Uploader Module (DUP) Functionality and Interaction.}
\label{fig13}
\end{figure}

The \texttt{Data Uploader Module (DUP)} is responsible for exfiltrating the data from the device, generated due to the execution of the C\&C command, to the server. It can upload any data stored on the device. This capability comes from the fact that it first accesses the memory contents and then writes the data as a byte stream to a file. After the file is created, and data is stored, it creates a connection to the server and uploads the data. The general functionality of DUP and its interaction with other components in the botnet is shown in figure \hyperref[fig13]{[Fig. 16]}.

\begin{figure}[!h]
\centering
\includegraphics[width=0.50\textwidth, height= 9cm]{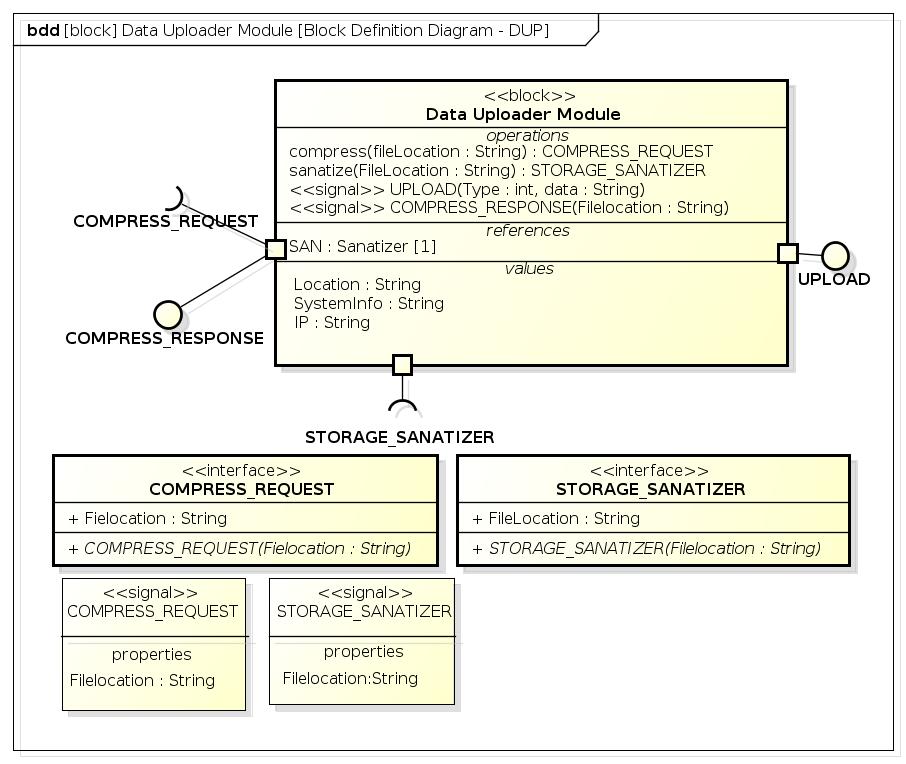}
\caption{Block Definition Diagram for Data Uploader Module (DUP).}
\label{fig14}
\end{figure}

The DUP provides an \textit{UPLOAD} interface, which is required by the Environment Manager. Data Uploader Module itself requires \textit{STORAGE\textunderscore SANATIZER} interface. As soon as, the file is uploaded to the specified server successfully, DUP calls the \texttt{Sanatizer Module}(SAN) to clean or sanitise the file from the device storage.  The SAN provides the \textit{STORAGE\textunderscore SANATIZER} interface. The DUP also requires another interface called \textit{COMPRESS\textunderscore REQUEST}, which is provided to it by the DC. Whenever DUP finds that the file to be uploaded is of considerable size, it calls the DC through the \textit{COMPRESS\textunderscore REQUEST} interface. The request contains the path to the file in the device storage. Then the DC compresses the contents of the file by using a pre-specified compression algorithm. After the compression process is over, the DC sends the path of the compressed file in storage to the DUP. The DUP copies the compressed file contents using the path and uploads the same to the server. Figure \hyperref[fig14]{[Fig. 17]} shows the Block Definition Diagram for the Data Uploader Module.

\subsubsection{Sanitizer Module (SAN)}
\label{san}
\begin{figure}[!h]
\centering
\includegraphics[width=0.45\textwidth, height= 7cm]{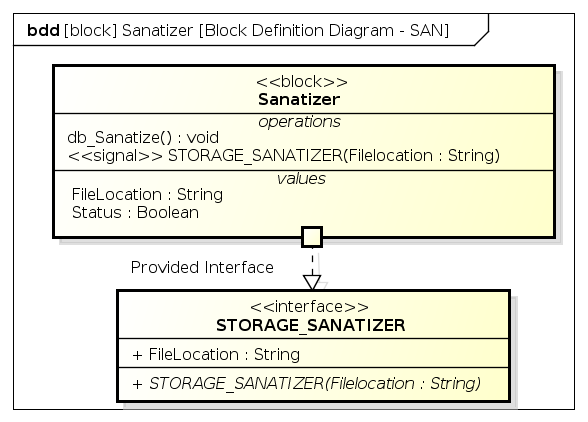}
\caption{Block Definition Diagram of Sanitizer Module (SAN).}
\label{fig8}
\end{figure}

The \texttt{Sanitizer module}, shown in figure \hyperref[fig8]{[Fig. 18]}, is used to sanitise the bot so that no traces are left behind for forensics when a command is executed. The SAN periodically deletes the command history and related information from the database. The commands that have an \textit{EXECUTED} status are completely removed from the database by the SAN. It provides the \textit{STORAGE\textunderscore SANATIZER} interface, which is required by the DUP. SAN also sanitises any secondary data that is generated because of the command execution on the device.

\section{CASE-STUDY (ANDROID FRAMEWORK)}
\label{case}
Since we have built an architecture of the model, now we will show how it can actually be implemented on an IoT device. We have a huge range of IoT devices available. From small scale IoT devices like home automation to sophisticated IoT devices like robotics on the assembly line,  from real-time IoT devices like those used in defence, aerospace, and healthcare to mobile IoT devices like personal digital assistants or mobile phones, anything can be used to explain the capabilities of a botnet. From the vast range of choices, we selected mobile phones, as they are most widely and globally used as far the current trend is considered and contain sensors which if exploited can be highly detrimental to user privacy\hyperref[R33]{[33]}. Since there are many variants of smartphones available in the market we have chosen, as our target, android platform because android maintained its position as the leading mobile operating system worldwide in October 2020, controlling the mobile OS market with a 72.92 percent share\hyperref[R34]{[34]}.

Avira (a multinational computer security software company) reports that the first quarter saw a full-force attack of the COVID-19 pandemic that left its mark on the Android ecosystem. Malware authors exploited this chaos along with the Android users’ need for information and protection. Further, they conclude that all types of malware, ranging from spyware and adware to more sophisticated banking trojans and SMS stealers, were distributed under the flag of COVID-19 related apps\hyperref[R35]{[35]}. Malwarebytes Labs reports that 2019 saw over 100 variants of stalkerware spyware with capabilities that allow it to be used to stalk or spy on someone else. That includes collecting the following data from someone else’s device without their informed consent: GPS location data, photos, emails, text messages, call logs, contacts lists, nonpublic social media activity, and more\hyperref[R36]{[36]}.\\
For these and other reasons, it is safe to say that the vast majority of mobile cyber-threats are targeting Android. So we think a good way to determine the behaviour and capabilities of the botnet, Android platform is a better choice.
\subsection{Implementation and Experimental Setup}
\label{I&E}
The implementation of both the Botnet Architecture and the Bot Framework is provided in APPENDIX I. Further, an experimental setup of the implementation is provided as well. 

\subsection{Evaluation}
\label{eval}
In the previous section we determined how the infection actually runs inside a bot, now we explore the findings and results by measuring the amount of network traffic transferred, power consumption, and execution status of different commands. 
\subsubsection{Network traffic}
\begin{figure}[!h]
\centering
\includegraphics[width=.50\textwidth, height= 5cm]{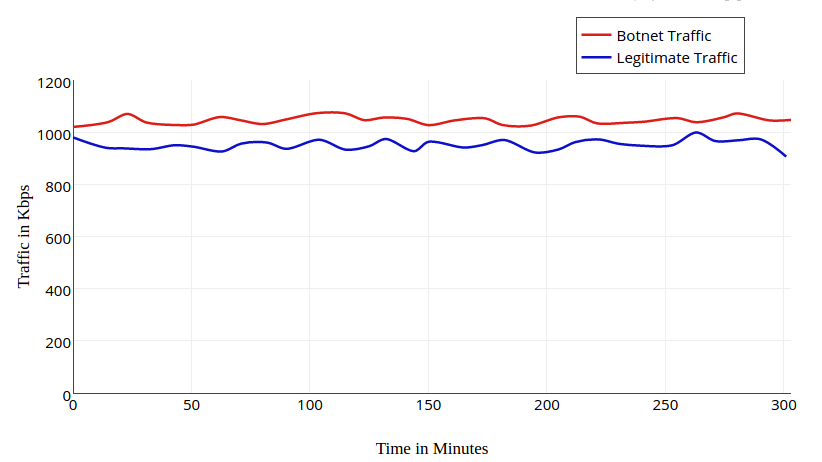}
\caption{Difference between Legitimate Traffic and Botnet Traffic}
\label{eval2}
\end{figure}

For this we analysed legitimate traffic of a device installed with five common applications like Whatsapp, Facebook, Gmail etc for about 5 hours. We installed a network traffic monitoring app on 5 devices. The users then used their phones during peak hours as normally they do. After the time period expired we used the average network traffic of these devices.\\
Then we analysed the traffic of an infected device with the same pre-installed applications for the same time frame i:e 5 hours. Since our bot does not hold a persistent connection with the C\&C  server we let it connect to the server at random instances of time. However, because the volume of the data to be uploaded, change as per the request, we, therefore, changed the commands, so that size of uploaded data is random also. This process was also repeated for 5 devices with varying versions of Android. Both the results are plotted on the graph for comparison  \hyperref[eval2]{[Fig. 25]}.
\\
To make the graph look smooth we plot the traces as the value of peak loads. For that, we take a data set of 10 entities from both groups with peak values only and plot the traces on the graph. The amount of legitimate traffic transferred over this particular time was 128.89\textit{Mbs} while the amount of traffic over the botnet was 142.34\textit{Mbs}, which is just a mere increment of about 10\%. This allows the botnet traffic to hide under the legitimate traffic, without being noticed by the user. With this little share of network traffic, we say the bot has a \textit{low bandwidth cost}.\\
\subsubsection{Power Consumption}
\begin{figure}[!h]
\centering
\includegraphics[width=.45\textwidth, height= 5cm]{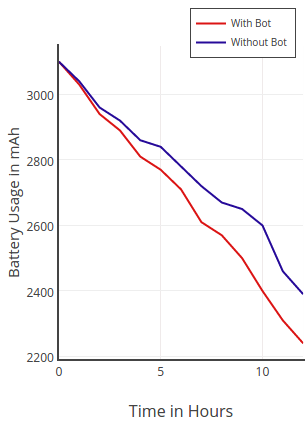}
\caption{Graph showing comparison of battery usages.\texttt{(Higher is Better)}}
\label{evalbat}
\end{figure}

In the following section, the battery use of bot is evaluated. We know that
low battery use is vital for the stealthiness of the bot. We compare results with and without using the bot. The result is shown below in the graph \hyperref[evalbat]{[Fig. 26]}.
\\
This experiment was done with a newly purchased phone analysing the battery consumption for 12 hours. During this comparison, the phone was not used at all, so the difference is expected to shrink if the screen was on for a reasonable amount of time, or if other power-consuming applications were used. When the phone was without a bot, the battery drained from 3100mAh to 2390mAh for 12 hours. The next day we again used the same phone with a full battery and installed bot, draining the battery from level 3100mAh to 2240mAh for the same time period, an overall decline of about 5\% which is marginally quite low. As also can be seen from the figure both plots look almost the same. Hence, it is extremely unlikely that the user discovers the bot solely based on its battery use. \\

%Table pre-plotted for the purpose of adjustment%
\begin{table*}[!]
\label{table}
%\twocolumn[{
\centering
\resizebox{\textwidth}{3cm}
{
\begin{tabular}{lllllllll}
%\toprule
\textbf{Device} & \textbf{Manufacturer}& \textbf{Model} & \textbf{OS Version} & \textbf{Sensors} & \textbf{Successful  Commands} &\textbf{Unsuccessful Commands} \\
%\midrule[0.08em]
Mobile & Samsung & On5 & Lollipop(21) / 5.1.1 & Camera, GPS, Mic & Capture,Location,Record & None  \\
\hline
Mobile & Samsung & Grand 2 & Kit Kat(19) / 4.4.2 & Camera, GPS, Mic & Location,Record & Capture  \\
\hline
Mobile & Xolo & Prime & Kit Kat(19) / 4.4.2 & Camera, GPS, Mic & Capture,Location,Record & None  \\
\hline
Mobile & Mi & Note Prime & Kit Kat(19) / 4.4.2 & Camera, GPS, Mic & None(only after notifying) & All  \\
\hline
Mobile & GioNee & add ur model & Lollipop(21) / 5.0.1 & Camera, GPS, Mic & Capture,Location,Record & None  \\
\hline
Mobile & Moto & E Ist Gen & Lollipop(21) / 5.0.1 & Camera, GPS, Mic & Capture,Location,Record & None  \\
\hline
Mobile & Micromax & Q340 & Lollipop(21) / 5.0 & Camera, GPS, Mic & Capture,Location,Record & None  \\
\hline
Mobile & Karbon & Titanium S12 & Kit Kat(19) / 4.4.2 & Camera, GPS, Mic & Capture,Location,Record & None  \\
\hline
Mobile & Samsung & Duos & Kit Kat(19) / 4.4.2 & Camera, GPS, Mic & Capture,Location,Record & None  \\
\hline
Mobile & Samsung & J7 & Lollipop (21) / 5.0.1 & Camera, GPS, Mic & Capture,Location,Record & None  \\
\hline
Mobile & Moto & G & Lollipop(21) / 5.0.1 & Camera, GPS, Mic & Capture,Location,Record & None  \\
\hline
Mobile & MI & 4 & Kit Kat(19) / 4.4.2 & Camera, GPS, Mic & None(only after notifying) & All  \\
\hline
Mobile & Panasonic & Eluga Mark 2 & Marshmallow(22) / 6.0.1 & Camera, GPS, Mic & Capture,Location,Record & None  \\
\hline
Mobile & HTC & Desire 628 & Lollipop(21) / 5.1.1 & Camera, GPS, Mic & Capture,Location,Record & None  \\
\hline
Mobile & Lenovo & Vibe K5 plus & Lollipop(21) / 5.1.1 & Camera, GPS, Mic & Capture,Location,Record & None  \\
\hline
Mobile & Samsung & Galaxy S8 & Noughat(24) / 7.1 & Camera, GPS, Mic & Capture,Location,Record & None  \\
\hline
Mobile & Oppo & F3 plus & Oreo(26) / 8.0 & Camera, GPS, Mic & Capture,Location,Record & None  \\
\hline
Mobile & Xiaomi & Note 8 Pro & Pie(28) / 9.0 & Camera, GPS, Mic & None(only after notifying) & All  \\
\hline
%\bottomrule
\end{tabular}
}
\caption{Table Showing Results of Commands on various devices.}
%}]
\end{table*}

\subsubsection{Defense Recommendations}
Several strategies to control the domain fluxing have been proposed as in \hyperref[R37]{[37]},\hyperref[R38]{[38]}, and \hyperref[R39]{[39]}. One technique that looks promising is using a blacklisting detection method but it is always difficult to build a blacklist and this method gives the attackers a time frame during which they can execute their attacks. We think a collaborative approach from ISP's, Domain Registration Authorities, and Researchers to tackle the domain fluxing strategy will prove to be a positive step to mitigate its effect. Techniques have been proposed to detect botnets based on the time they generate traffic like in \hyperref[R31]{[31]} and \hyperref[R32]{[32]} but a methodology should also be formalised to detect traffic during peak hours of usage. During the experiments, we found the security mechanism of various devices is weak because they allow permission to use sensors easily and without any effort. For that, it will be good if the user is notified when an application tries to use the sensors of a device, a feature which we found on many Xiaomi phones. Another important concern is the spam SMS, and the applications should be refrained to read SMS's from untrusted sources, as we showed how commands are embedded inside them. We advocate to study botnets, new theories and strategies are to be devised as in \hyperref[R40]{[40]}. Finally, during our discussion botnet writers may have developed new strategies for detection. It will be useful to use machine learning algorithms to predict their possible strategies.\\
\subsubsection{Enumeration of Commands Executed}
\label{enumeration}
This enumeration provided us with an analysis of the permission mechanism of Android devices. We enumerated each command on every device and the Table \hyperref[table]{[Table 1]} depicts our findings.
%\iffalse

%\fi

While enumerating the bot we found that most of the devices after installation gave the infected app full permissions to all the sensors, due to this lame approach the bot started working on the devices successfully and easily.
It's worth mentioning here that mobile phones like Samsung, Gionee, Motorola all provided full access to the sensors after installation but this behaviour was not seen in Xiaomi phones.
The Xiaomi device users provided a tough challenge to our bot because the Security Center of the Xiaomi phones did not provide full access to the hardware after the app installation. At the time of using any application on these phones, they notify the user if the application tries to access the hardware, and only if the user allows the application is granted access.\\

\section{CONCLUSION}
\label{conclusion}
In this paper, we provided a mechanism to exploit the security of an IoT device. The Botnet we proposed covers an ample and capacious range of IoT devices incorporated with various sensors. Apparently, there is no silver bullet for security and none of the systems claims to be 100\% secure. Although recent work and research in this field yielded positive results yet the challenges imposed by the process of securing emerging environments or networks of IoT devices compel us to study the problem again. The "component-based framework" is used to easily create a bot so that the adverse effects of a bot on the device are studied in detail and later some countermeasures are provided to nullify or dwindle its calamitous and detrimental effects (offence is the best defence). Apart from this, a full-fledged Botnet can be contrived to study its efficacy at the network level and counteractants be proposed to mitigate its compelling and dominant imprint. We need to have intelligent malware scanners, and firewalls for the IoT devices which can help in making it difficult to get the device infected and controlled remotely. Ultimately it is the same never-ending battle between the bad guys, who are up to mischief, and the security researchers, who try to make the system secure.

\section{COMPLIANCE WITH ETHICAL STANDARDS}
\label{compliance}
\textbf{\textit{Conflict of interest : }}The authors declare that they have no confict of interest.\\
\textbf{\textit{Ethical approval : }} This article does not contain any studies with human
participants or animals performed by any of the authors.\\
\textbf{\textit{Informed consent : }} None.\\
%\section*{References}

%\bibliography{mybibfile}

\begin{thebibliography}{37}
	\bibliographystyle{alpha}
	%Sri Parameswaran, Tilman Wolf, “Embedded systems security - an overview”, Design Autom. for Emb. Sys. 2008.
	\bibitem{RefJ}
	Ravi S, Raghunathan A, Chakradhar S (2004) Tamper resistance mechanisms for secure, embedded
	systems. In: 17th international conference on VLSI design, January 2004
	\label{R1}
	\bibitem{RefJ}
	O. Kommerling and M. G. Kuhn, “Design principles for tamper-resistant smartcard
	processors,” in Proc. USENIX Wkshp. on Smartcard Technology (Smartcard ’99), pp. 9–20, May 1999.
	\label{R2}
	\bibitem{RefJ}
	"Smart Card Handbook", John Wiley and Sons.
	\label{R3}
	\bibitem{RefJ}
	Shuang Zhao, Patrick P.C.Lee, John C.S. Lui, Xiaohong Guan, Xiaobo Ma, Jing Tao, "Cloud-based Push-Styled Mobile Botnets: A Case Study of Exploiting the Cloud to Device Messaging Service", ACSAC '12 Proceedings of the 28th Annual Computer Security Applications Conference,
	Pages 119-128, 2012.
	\label{R4}
	\bibitem{RefJ}
	Alberto Ornaghi, Marco Valleri, "Man in the middle attacks", Blackhat Conference Europe 2003, \url{https://www.blackhat.com/presentations/bh-europe-03/bh-europe-03-valleri.pdf}
	\label{R5}
	\bibitem{RefJ}
	Ulrike Meyer, Sussane Wetzal "A Man-in-the-Middle Attack on UMTS", WiSe’04, October 1, 2004, Philadelphia, Pennsylvania, USA.
	\label{R6}
	\bibitem{RefJ}
	Common Vulnerabilities and Exposures. \url{https://cve.mitre.org/}
	\label{R7}
	\bibitem{RefJ}
	E. Chien and P. Szor, "Blended attack exploits, vulnerabilities, and buffer overflow
	techniques in computer viruses.", Symantec White Paper,  \url{https://www.symantec.com/avcenter/reference/blended.attacks.pdf}.
	\label{R8}
	\bibitem{RefJ}
	Hagai Bar-El, "Introduction to Side Channel Attacks." Discretix White Paper, 
	\url{http://gauss.ececs.uc.edu/Courses/c653/lectures/SideC/intro.pdf}
	\label{R9}
	\bibitem{RefJ}
	P. Kocher, J. Jaffe, and B. Jun, Introduction to differential power analysis
	and related attacks. (http://www.cryptography.com/resources/
	whitepapers/).
	\label{R10}
	\bibitem{RefJ}
	T. S. Messerges, E. A. Dabbish, and R. H. Sloan, “Examining Smart-Card Security
	under the Threat of Power Analysis Attacks,” IEEE Trans. Comput., vol. 51,
	pp. 541–552, May 2002.
	\label{R11}
	\bibitem{RefJ}
	P. Kocher, J. Jaffe, and B. Jun, “Differential Power Analysis,” Advances in Cryptology
	– CRYPTO’99, Springer-Verlag Lecture Notes in Computer Science, vol. 1666,
	pp. 388–397, 1999.
	\label{R12}
	\bibitem{RefJ}
	D. Brumley and D. Boneh, “Remote Timing Attacks Are Practical ,” in Proc. 12th
	USENIX Security Symp., pp. 1–14, Aug. 2003.
	\label{R13}
	\bibitem{RefJ}
	P. C. Kocher, “Timing attacks on implementations of Diffie-Hellman, RSA, DSS,
	and other systems,” Advances in Cryptology – CRYPTO’96, Springer-Verlag Lecture
	Notes in Computer Science, vol. 1109, pp. 104–113, 1996.
	\label{R14}
	\bibitem{RefJ}
	J.F. Dhem, F. Koeune, P.A. Leroux, P. Mestre, J.J. Quisquater, and J.L. Willems,
	“A practical implementation of the timing attack,” in Proc. Third Working Conf.
	Smart Card Research and Advanced Applications, pp. 167–182, Sept. 1998.
	\label{R15}
	\bibitem{RefJ}
	Boneh D, DeMillo RA, Lipton RJ, "On the importance of eliminating errors in cryptographic
	computations.", J Cryptol 14(2):101–119, 2001
	\label{R16}
	\bibitem{RefJ}
	Boneh D, DeMillo RA, Lipton RJ, "On the importance of checking cryptographic protocols for
	faults.", In: Lecture notes in computer science, vol 1233. Springer, Berlin, pp 37–51, 1997
	\label{R17}
	\bibitem{RefJ}
	D. Agrawal, B. Archambeault, J.R. Rao, P. Rohatgi, "The EM Side–Channel(s).", CHES
	2002,LNCS 2523,pp.29-45,2003.
	\label{R18}
	\bibitem{RefJ}
	K. Gandolfi,C.Mourte,F. Olivier, "Electromagnetic Analysis: Concrete Results.", CHES
	2001,LNCS 2162,pp.251-261, 2001.
	\label{R19}
	\bibitem{RefJ}
	J.J. Quisquater, D. Samyde, "Electromagnetic analysis (EMA): measures and countermeasures
	for smart cards.", E-smart 2001,LNCS 2140,pp.200–210,2001.
	\label{R20}
	\bibitem{RefJ}
	M.G. Kuhn, R.J. Anderson, "Soft tempest: hidden data transmission using
	electromagnetic emanations.", Information Hiding 1998,LNCS 1525,pp.124-142,1998.
	\label{R21}
	\bibitem{RefJ}
	E.D. Mulder, P. Buysschaert, S.Börs, P. Delmotte, B. Preleel, G. Vandenbosch, I.
	Verbauwhede, "Electromagnetic Analysis Attack on a FPGA Implementation of an Elliptic Curve
	Cryptosystem.", Katholieke Universiteit Leuven, Leuven, Belgium.January 2005. 
	\label{R22}
	\bibitem{RefJ}
	"Massive Botnet Attack Used More Than 400,000 IoT Devices", July 26, 2019, \url{https://www.bankinfosecurity.com/massive-botnet-attack-used-more-than-400000-iot-devices-a-12841/}
	\label{R23}
	\bibitem{RefJ}
	"Spamhaus Botnet Threat Report 2019", \url{https://www.deteque.com/app/uploads/2019/02/Spamhaus-Botnet-Threat-Report-2019.pdf}
	\label{R24}
	\bibitem{RefJ}
	"Vigilante botnet highlights woeful state of embedded device security", Oct 5, 2015, \url{https://securityledger.com/2015/10/vigilante-botnet-highlights-woeful-state-of-embedded-device-security/}
	\label{R25}
	\bibitem{RefJ}
	"Thousands of hacked CCTV devices used in DDoS attacks", June 28, 2016. \url{http://www.pcworld.com/article/3089346/security/thousands-of-hacked-cctv-devices-used-in-ddos-attacks.html}
	\label{R26}
	\bibitem{RefJ}
	"Over 100 DDoS botnets built using Linux malware for embedded devices", June 30, 2016, \url{http://www.csoonline.com/article/3090161/security/over-100-ddos-botnets-built-using-linux-malware-for-embedded-devices.html}
	\label{R27}
	\bibitem{RefJ}
	"Botnets of Embedded Devices Are Brute-Forcing Telnet Ports" Sep 7 2016",  \url{http://news.softpedia.com/news/botnets-of-embedded-devices-are-trying-to-brute-force-telnet-ports-508050.shtml}
	\label{R28}
	\bibitem{RefJ}
	P. Traynor, M. Lin, M. Ongtang, V. Rao, T. Jaeger, T. La Porta and P.McDaniel, “On Cellular Botnets: Measuring the Impact of Malicious Devices on a Cellular Network Core,” in ACM Conference on Computer and Communications Security (CCS), November 2009.
	\label{R29}
	\bibitem{RefJ}
	R. Villamarín-Salomón and J.C. Brustoloni, "Identifying botnets using anomaly detection techniques applied to DNS traffic.", In Consumer Communications and Networking Conference, 2008. CCNC 2008. 5th IEEE, pages 476–481. IEEE, 2008.
	\label{R30}
	\bibitem{RefJ}
	B. AsSadhan, J.M.F. Moura, D. Lapsley, C. Jones, and W.T. Strayer. "Detecting botnets using command and control traffic.", In Network Computing and Applications, 2009. NCA 2009. Eighth IEEE International Symposium on, pages 156–162. IEEE, 2009.
	\label{R31}
	\bibitem{RefJ}
	F. Giroire, J. Chandrashekar, N. Taft, E. Schooler, and D. Papagiannaki. "Exploiting temporal persistence to detect covert botnet channels.", In Recent Advances in Intrusion Detection, pages 326–345. Springer, 2009.
	\label{R32}
	\bibitem{RefJ}
	"Gartner Says Worldwide Smartphone Sales Will Grow 3\% in 2020",
	\url{https://www.gartner.com/en/newsroom/press-releases/2020-01-28-gartner-says-worldwide-smartphone-sales-will-grow-3--}
	\label{R33}
	\bibitem{RefJ}
	"Mobile operating systems' market share worldwide from January 2012 to October 2020",
	\url{https://www.statista.com/statistics/272698/global-market-share-held-by-mobile-operating-systems-since-2009/}
	\label{R34}
	\bibitem{RefJ}
	"Malware Threat Report: Q2 2020 Statistics and Trends",
	\url{https://www.avira.com/en/blog/malware-threat-report-q2-2020-statistics-and-trends}
	\label{R35}
	\bibitem{RefJ}
	"2020 State of Malware Report",
	\url{https://resources.malwarebytes.com/files/2020/02/2020_State-of-Malware-Report.pdf}
	\label{R36}

	\bibitem{RefJ}
	Sandeep Yadav, Ashwath Kumar Krishna Reddy, A. L. Narasimha Reddy, Supranamaya Ranjan, "Detecting Algorithmically Generated Domain-Flux Attacks With DNS Traffic Analysis", IEEE/ACM Transactions on Networking ( Volume: 20, Issue: 5, Oct. 2012 ), Pages 1663 - 1677
	\label{R37}
	\bibitem{RefJ}
	M. Antonakakis, R. Perdisci, Y. Nadji, N. Vasiloglou, S. Abu-Nimeh, W. Lee, and D. Dagon, "From throw-away traffic to bots: Detecting the rise of DGA-based malware", in Proceedings of the 21st USENIX Security Symposium, Bellevue, WA, USA, pp. 24-40, August 2012.
	\label{R38}
	\bibitem{RefJ}
	L. Bilge, E. Kirda, C. Kruegel, and M. Balduzzi, "Exposure: Finding malicious domains using passive DNS analysis", in Proceedings of the Network and Distributed System Security Symposium (NDSS), San Diego, CA, USA, February 2011.
	\label{R39}
	\bibitem{RefJ}
	Guining Geng, Guoai Xu, Miao Zhang and Yanhui Guo, "The Design of SMS Based Heterogeneous
	Mobile Botnet.", Journal of Computers, Vol. 7, No. 1, pages 235-243, 2012.
	\label{R40} 
	\bibitem{RefJ}
	Yuanyuan Zeng, Kang G.Shin, Xin Hu, "Design of SMS Commanded-and-Controlled and P2P-Structured Mobile Botnets.", WISEC 12 Proceedings of the fifth ACM conference on Security and Privacy in Wireless and Mobile Networks, pages 137, 2012. 
	\label{R41}
	%P. Van Mieghem, Graph Spectra for Complex Networks. Cambridge press, 2011.
	%\label{B25}
	
\end{thebibliography}

\section{APPENDIX I}
\label{case}
\subsection{Deployment of the overall Botnet}
\label{DeployNA}

In our implementation, we used Android mobile devices as bots, online servers as C\&C servers, and Android devices as botmaster also. The communication in the network occurs as was described in the model \hyperref[BBArch]{[Fig. 1]} with a little refinement. For communication, we used a hybrid network, in which we used two set of communication lines Internet and SMS. The potential advantages of using SMS as a communication medium has been analysed in \hyperref[R41]{[41,4]}. The SMS costs incurred, when SMS channel is used for communication from the bot to botmaster, may alert a device owner about the presence of botnet payload activity, and in turn, may result in the disinfection of the device. To overcome this drawback, we have used SMS channel as a one-sided channel in the network. The SMS messages are used as commands for the bot. Once the command is delivered through the SMS channel, it will be executed and the bot will wait for other channels of communication to be active to send the requested information back to the botmaster. Since, SMSs are usually sent from the botmaster, which itself is a mobile phone with a unique identity, a point of concern that rises here is that it may lead to the disclosure of the identity of the botmaster. To remedy this concern, the SMSs can be sent using currently available online SMS gateways, or by setting up one for the purpose.

\subsection{Deployment of the Bot Framework}
\label{Deploy Framework}
To demonstrate the functioning of bot we exploited the sensors of the Android phone without letting the users know e:g details from sensors like GPS, Camera, Mic are captured by the bot and then uploaded to the botmaster. We choose these sensors because they are the most widely and frequently used but with a little more effort any other sensors of the device are to be exploited to retrieve the information. GPS is used to track the location of the bot, Camera captures the pictures stealthily, and Mic is used to record audio, and calls from the bot.\\

In implementing the functionality we use the framework of the bot given earlier and for that, we extend some modules. The \texttt{Receiver Module} (REC) has been split into two modules namely SMS Receiving Module (SRM) and Internet Reciever Module(IR). Also, the \texttt{Device Status Generator Module} (DSG) has been renamed to Phone Status Generator Module(PSG). All the other modules stay unmodified including the Bot Module. Now we explain the functionality of each module.
\subsubsection{The Bot}
\label{CSBot}
This unit is actually identifying a device in a network. It has two value types: \textit{Mobile Number} and \textit{Device\textunderscore id}. The \textit{Mobile Number} specifies the device that is infected, and the \textit{Device\textunderscore id} is the identity of the bot in the network system.
Once installed on an Android device the Bot will try to extract the mobile number of the device. Later during the registration process, the bot is assigned an id.
\subsubsection{Receiver}
\label{CSRec}
Although the Android framework is designed such that it is flexible enough to use any of the communication channels like Internet, SMS, WiFi, Bluetooth, Infrared etc, we have implemented the SMS and the internet channels for testing purposes. The reason behind choosing these two channels of communication is their frequent use in the devices selected for the case study. As such, we inherit two sub-modules from the Reciever Module: SMS Receiving Module (SRM) and Internet Receiver Module (IR). The Block Definition Diagram of the Receiver (REC) Module is given in figure \hyperref[fig4]{[Fig. 20]}.

\begin{figure}[!h]
\centering
\includegraphics[width=0.50\textwidth, height= 7cm]{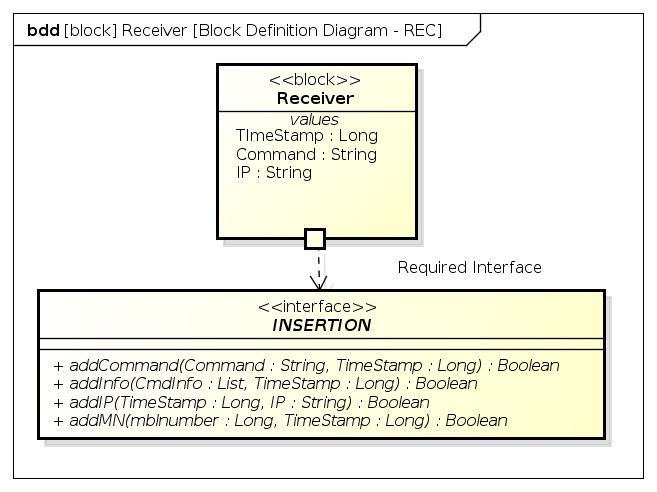}
\caption{Block Definition Diagram of Receiver (REC) Module.}
\label{fig4}
\end{figure}

\begin{figure}[!h]
\centering
\includegraphics[width=0.5\textwidth, height= 7cm]{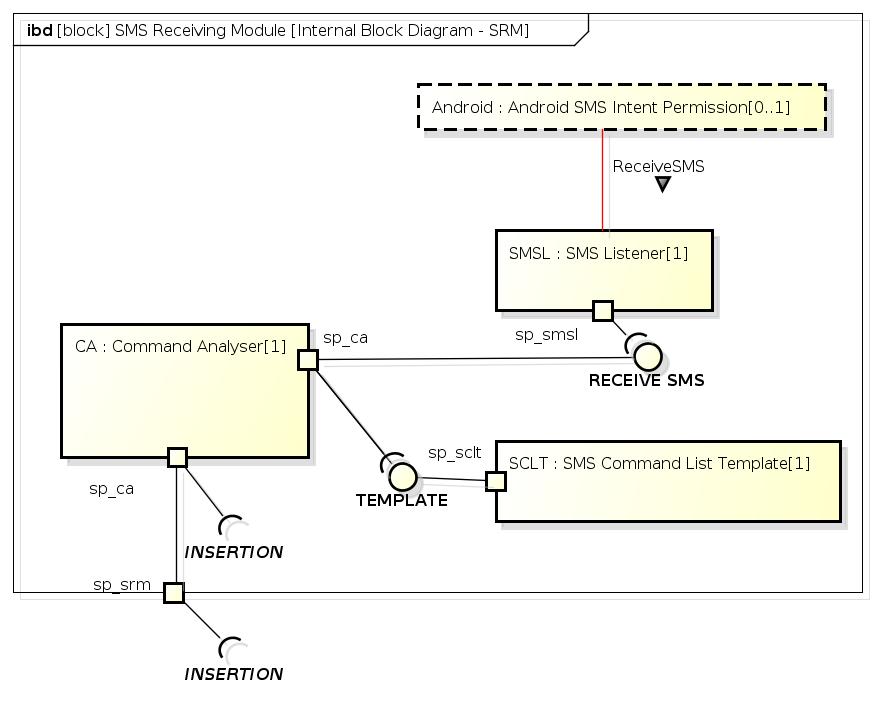}
\caption{Internal Block Diagram of SRM}
\label{SRM}
\end{figure}

\begin{figure}[!h]
\centering
\includegraphics[width=0.70\columnwidth,height = 2.5cm]{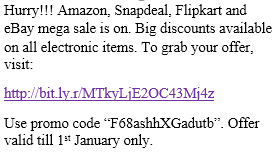}
\caption{CAPTURE Command SMS Template.}
\label{fig29}
\end{figure}

\subparagraph{SMS Receiving Module(SRM):}
\label{srm}
The SMS Receiving Module (SRM) is a sub-type of Receiver (REC) Module, which enables the Android device to use SMS communication channel for C\&C purposes. On an Android device, to make messages look like as usual spam messages sent by telemarketing agencies, many message templates have been incorporated. The botmaster can choose any of the templates and craft a new SMS from it.   \\
Even though SMS as a C\&C channel has certain drawbacks. There are a few advantages associated with using SMS for our purpose. The first advantage is its ubiquity as a communication medium in mobile phones. It is most widely used data application on the planet when considering mobile phones. The second advantage offered by SMS as a C\&C channel is its almost persistent availability. When a mobile phone is turned on the SMS service is activated automatically. The third advantage is that when SMS is used as a C\&C channel, to communicate with the bot no additional services are required to be active on the device. The fourth advantage of SMS-based C\&C channel is that it can accommodate offline bots. For example, if a mobile phone is turned off or has a poor signal reception at the time a C\&C SMS was sent to it, the SMS containing the C\&C information will be stored in a service centre provided by the service provider. The SMS will be delivered to the device as soon as it is turned on or reaches reception zone. \\
The SMS Receiving Module (SRM) itself is not a concrete individual module, but constitutes three sub-modules, as shown in internal block diagram \hyperref[SRM]{[Fig. 21]}. Since SRM is a composite module, so its functioning is determined by its constituent internal blocks or sub-modules.The SRM can be decomposed into three sub-modules or sub-blocks. It also shows that SRM requires an additional external reference entity, named as \textit{Android SMS Intent Permission}. In addition to these components, SRM also requires an interface named \textit{Insertion Interface}.  We discuss these sub-modules in some detail to understand what functionality each of these sub-modules performs and how they interact within the SRM.\\
\label{smsl}
\textit{a. Listener Module (SMSL):} SMSL is an important component of SRM. It gets invoked when SMS is received on the device. The SMSL requires a receiver interface named as \textit{RECEIVE\textunderscore SMS} interface. When SMSL is invoked on the \textit{SMS\textunderscore RECEIVED} event, it uses this interface to pass on the information to other components of the bot. The component to which the information is passed is Command Analyzer (CA). The \textit{RECEIVE\textunderscore SMS} interface is provided to SMSL by the Command Analyzer module.\\
 An old trick to put SMSL to work was to get it registered as a listener to the \textit{SMS\textunderscore RECEIVED} event of the device. When a new SMS arrived on the device, all the registered listeners were notified about the event. The notifying process worked as per a pre-determined priority assigned to different listeners. So, it started notifying with the listener which had the highest priority. The priority scheme was obfuscated so that it assigned the highest priority to the SMSL. By doing so, the bot was able to get the copy of SMS before any other listener. After this, the bot checks the SMS contents to verify whether it was meant for it or not. If the SMS was meant for the bot or it was a command-and-control SMS, the bot stops its further propagation among other listeners by using the \textit{abortBroadcast()}. So, the contents of the SMS sent by the botmaster remains hidden from the user.Although a good trick, but with the consequent security updates done by manufacturers of such devices, the feature of an aborting broadcast of SMS to other listeners works only with devices which support Ordered Broadcasts.   \\
To overcome the hurdles created by these security upgrades in command-and-control propagation through SMS, the authors took advantage of SMS spamming. Since SMS spamming is common and an average user receives spam SMSs. To make the bot more efficient, the authors replaced the aforementioned old trick by using the traditional spamming technique.  The botmaster encodes commands for the bot into spam-like messages so that if a user checks the contents of the SMS he treats it as a spam SMS.\\
\label{sclt}
\textit{b. SMS Command List Template Module (SCLT):} SMS allows a fixed length message, usually 160 characters, to be sent as a single message. For this reason, the commands in this bot have been designed to be as much concise as possible. For example, the \textit{CAPTURE} message instructs a bot to take an image and upload it to the specified server. A visual representation of how the \textit{CAPTURE} command template looks inside the SMS sent by the botmaster is given by figure \hyperref[fig29]{[Fig. 22]}.

As is obvious from figure \hyperref[fig29]{[Fig. 22]}, the SMS received by the user does not display the actual contents of the message. This is because in this botnet each message is disguised before it is sent to the target. This is done by mapping each message template to a specific spam template. Additional information associated with a particular command is also encoded suitably into the message. \\
After the message is received by the bot, the message is decoded first. This is done by remapping of SMS templates to their corresponding commands, which is done by SCLT. The mapping is done by using a pre-specified hash function. The SCLT stores a list of hash values, wherein each hash value corresponds to a particular message template, and thus represents a particular command for the bot. When a command-and-control SMS is received its hash value is searched in the list of hash values stored in SCLT. If a match is found, the corresponding command is returned to the bot for execution.    \\
The SCLT, in addition to performing mapping, provides additional information that may be associated with the SMS. For instance, the SMS template given in figure \hyperref[fig29]{[Fig. 22]} represents the \textit{CAPTURE} command. The \textit{CAPTURE} may have associated with it additional arguments like the IP address of the server to which the image is to be uploaded after the capture operation. Let us suppose, the server IP address is something like 192.168.xx.x. To disguise this IP address, it is encrypted using Base64 encryption, and then embedded into the SMS. The Base64 encryption of 192.168.xx.x using UTF-8 character set produces MTkyLjE2OC43Mi4z.  \\
To make detection harder, one command message can correspond to different spam templates, and the templates can be updated periodically. If a user deletes the message, it will not cause any problem because the command will be executed as soon as the message is received by the device. In addition to performing these functions, SCLT provides the \textit{TEMPLATE} interface to the Command Analyzer module.

\label{ca}

\textit{c.Command Analyzer:} The Command Analyzer module of SRM, determines the command an SMS represents, and then how it is to be executed. To determine the command, it requires the contents of an incoming SMS along with the additional information from SCLT. The CA gets the hash value of SMS content, which it verifies with the help of SCLT. The SCLT also helps CA to determine the additional information associated with an SMS. Finally, it inserts all the details related to the SMS into the local database of the bot using the \textit{INSERTION} interface. In case, the SMS is not a C\&C SMS the Command Analyzer discards it completely.\\
An interface, like a block or module, is an element of the definition. It defines a set of operations and acts as a behavioural contract between the receivers and the providers. An interface actually defines rules for communication between two interfacing modules. SMS Receiving Module has three interfaces associated with it.
\subparagraph{Internet Receiver Module (IR)}
\label{ir}
The Internet Receiving (IR) module is another sub-module of the Receiver Module (RM). The IR module enables the device to use the wired or wireless internet as the command-and-control channel. So, for the botnet to send and receive data over the internet, it must use the \textit{Internet Receiver (IR)}. The Internet Receiver (IR) module is important because it is the main channel through which the proposed botnet connects the various components of the system like botmaster, command-and-control servers, and bots. The Internet Receiver helps in overcoming the challenges faced while using SMS as a command-and-control channel. It enables the botnet to establish a duplex link between the botmaster and the bots, as opposed to the simplex link provided by the SMS channel.   \\
The bandwidth of the internet channel is much greater than that of the SMS channel. The IR channel network interface is the main network interface which we use for the implementation of the proposed botnet. Since it is the main receiving interface of the botnet, it needs to be secured efficiently.   \\

The Internet Receiver (IR) module deals with three types of values: DEVICE\textunderscore ID, COMMAND\textunderscore TYPE, and TIMESTAMP. The number of operations that IR supports, depends upon the platform for which this framework is implemented. The IR module supports three operations: \textit{getDeviceId()}, \textit{getIPaddress()}, and \textit{HttpPostRequest(Url: String, RequestParams: List)}. \\

The \textit{getDeviceId()} extracts the target device's id, which is usually unique to every Android phone. This id is used to identify different phones on the botnet. The \textit{getIPaddress()} extracts the IP address currently assigned to the device. This IP address is used to establish an internet connection with the target device by the botmaster and C\&C servers. The \textit{HttpPostRequest()} sends the request to the server. \textit{HttpPostRequest()} takes arguments which include URL of the remote server and the various parameters that the server side scripts need to process the bot's requests.   \\

The IR module uses the \textit{Download Command Request (DCR)} to get C\&C information for the bot. The IR module periodically polls the server with DCR requests. When the server receives the request, it checks whether any command has been published by the botmaster for the bot's device id. If there is a command, it sends the response to the bot. This response contains the command, timestamp, parameters and the IP address. IR then concatenates the Device\textunderscore  Id and timestamp just received into one Unique\textunderscore ID. This information received by the bot is inserted into the bot database using the \textit{INSERTION} interface. Instead of using the server-side \textit{PUSH} mechanism to push commands to the bot, the IR uses \textit{PULL} mechanism for obtaining command-and-control information from the server. This makes the system stealthy because no commands are broadcasted, instead, only those commands are sent which are \textit{PULL} requested by any bot.

\subsubsection{Command and Control Collector}
\label{CSccc}
The Command and Control Collector (CCC) unit build a special SQLite database with a single table having the following attributes.
\begin{itemize}
\item Command.
\item Timestamp.
\item Parameters.
\item Status.
\end{itemize}

The Command and Control Collector here provides two interfaces. Firstly a command insertion interface that is used to add commands to the database, secondly a command dispatcher interface through which commands are retrieved from the database.
\paragraph{Command Insertion:} This unit is responsible for inserting and updating the contents of the database.It uses two methods like \textit{add()} and \textit{update()}. As a method \textit{add()} accepts three arguments \textit{Command}, \textit{Timestamp }and \textit{Parameters} while as update() also accepts three arguments except that the third argument is \textit{Status}. When a new entry is inserted using add(), the status field is automatically updated to "PENDING". Later we can update its contents using the update method.
\paragraph{Command Dispatcher:} The Command Dispatcher unit helps in retrieving the commands. It has one method \textit{GetPendingCommands()} and takes a single argument \textit{Status}. This argument is used to retrieve only the desired commands e:g if the value of \textit{Status} field is \textit{"PENDING"} then only the commands with this property are returned by the method.
\subsubsection{Environment Manager}
\label{CSem}
Firstly this unit checks whether the battery is feasible enough to run the bot and if it determines that the battery level is below the threshold it asks the bot to sleep until the battery reaches the desired level. In our implementation the bot then needs to register the device, it does so by first retrieving the information like \textit{device\textunderscore id, IP address, username, mobile no} e.t.c of the device and then sends the same information to the server. It then waits for an acknowledgement and after receiving an acknowledgement it creates a persistent variable to let the bot know in future that the current device has been registered. \\

After the registration is done the Environment manager continuously polls the server through Internet Reciever unit to download a command, if present. Later it checks the database for any pending or half-executed commands. If a command with status value \textit{PENDING} is found the Environment Manager tries to execute the command by first checking the status of sensors used by the command via Phone Status Generator and if it receives a clearance it executes the command. Once a command is executed to its entirety the status of that particular command is set to \textit{HALF-EXECUTED}. Then if the Environment manager finds a suitable communication link it tries to upload the data to the server and if it succeeds it sets the status of that particular command to \textit{EXECUTED}. It also invokes the Sanitizer via File Upload Unit to remove the data traces from the bot. 
\subsubsection{Phone Status Generator}
\label{CSpsg}
To use various resources of the device like a camera, microphone, GPS and battery, the Environment Manager first requires knowing about the capabilities and status of these resources. The status of these resources is provided by the Phone Status Generator, thereby allowing the Environment Manager make a decision to execute a command only when that particular resource is available. In this way, it helps save computation time. Information of these resources is provided by the Device Heartbeat Unit. In addition to above, the bot uses information received from the battery sensor to control the functioning of the bot to overcome the Battery Drainage problem. The bot suspends itself if the battery sensor indicates that the battery level is below the threshold and is not plugged in. Similarly, the bot uses the information from the memory sensor to determine if it is feasible to store new data on the device storage. Details of these sensors are provided by the Sensor Unit. 

\subsubsection{Device Heartbeat}
\label{CSdhb}
The Device Heartbeat (DHB) unit extracts information about the primary and the secondary storages of the device. It provides the information about the total and available capacity of the device RAM. It also provides the Phone Status Generator with the information like total and available, internal and external device storage capacity. It is also responsible for providing the information regarding the device battery like battery level \textit{(0 to 100)} and charging status \textit{(True, false)}.\\

The Device Heartbeat module is composed of methods which are directly called by the phone Status Generator. The memory status related input parameters are received only by the \textit{getMemorySize()} method, which then takes a decision to return the requested information. It takes as input either \textit{getFreeSize}, \textit{getUsedSize}, or \textit{getTotalSize}, and returns the requested information to the calling module. The battery status related input parameters are received by the \textit{BatteryStatus()} method. The battery status related information is acquired by registering a BroadcastReceiver of intent type, which is called \textit{ACTION\textunderscore BATTERY\textunderscore CHANGED}. The bot supports two properties of the battery. First, the \textit{BATTERY\textunderscore LEVEL} for which the Environment Manager sets a threshold which determines whether the level of the battery is feasible to perform a particular operation. Second, \textit{BATTERY\textunderscore STATUS\textunderscore CHARGING}, which determines if the battery is currently charging, reducing the criticality of the low \textit{BATTERY\textunderscore LEVEL} constraint.

\subsubsection{Sensor Unit}
\label{sm}
\begin{figure}[!h]
\centering
\includegraphics[width=.50\textwidth, height= 10cm]{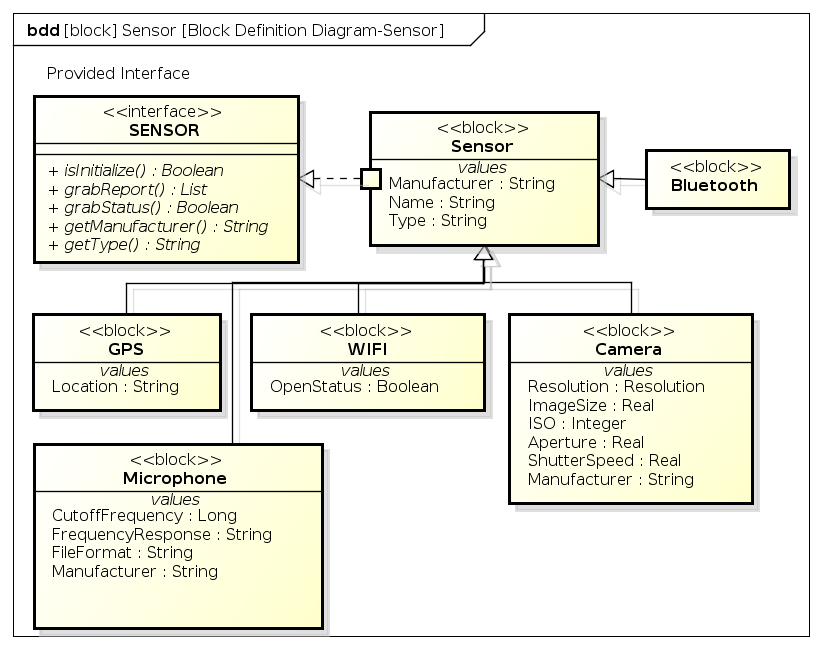}
\caption{Block Definition Diagram of Sensor Unit}
\label{figsm}
\end{figure}

The Sensor Unit is bifurcated into Connectivity Sensors like GPRS, Wi-Fi, and Bluetooth etc, and phone sensors like Microphone, GPS, and Camera. Figure \hyperref[figsm]{[Fig. 23]} shows the generic architecture and the various attributes and methods implemented by different sensors constituting the Sensor Module.
\\
The connectivity sensor \textit{(NetworkInfo)} is responsible for providing information mainly related to the various network attributes. The information provided by it is usually related to available communication channels like Wi-Fi, Bluetooth, and the internet. It provides details like service operator name, SIM serial number, and Bluetooth serial. It uses two types of services to acquire this information from the device. It uses \textit{CONNECTIVITY\textunderscore SERVICE} provided by the connectivity manager to get the status of the sensors like Wi\textunderscore Fi, Bluetooth, and the internet. It also uses \textit{TELEPHONY\textunderscore SERVICE} provided by the telephony manager to acquire the information related to the SIM.  \\
The camera sensor \textit{(GrabCameraDetail)} first determines the available number of cameras in the device. Once determined, it loops over that number to find the details of each camera. The details provided by this sensor include focal length, jpeg quality, shutter speed, and preview size.   \\
The microphone sensor \textit{(GrabMicDetail)} creates an object of the media recorder and then accesses the status of the microphone. It provides details like cutoff frequency, frequency response, file format, and encoding type.  \\
The GPS sensor (\textit{GrabGPSDetails}) creates an object of the location manager and then uses the \textit{LOCATION\textunderscore SERVICE} provided by the location manager to determine the status of GPS.
\subsubsection{Command Execution Unit}
\label{CSceu}
\begin{figure}[!h]
\centering
\includegraphics[width=.50\textwidth, height= 10cm]{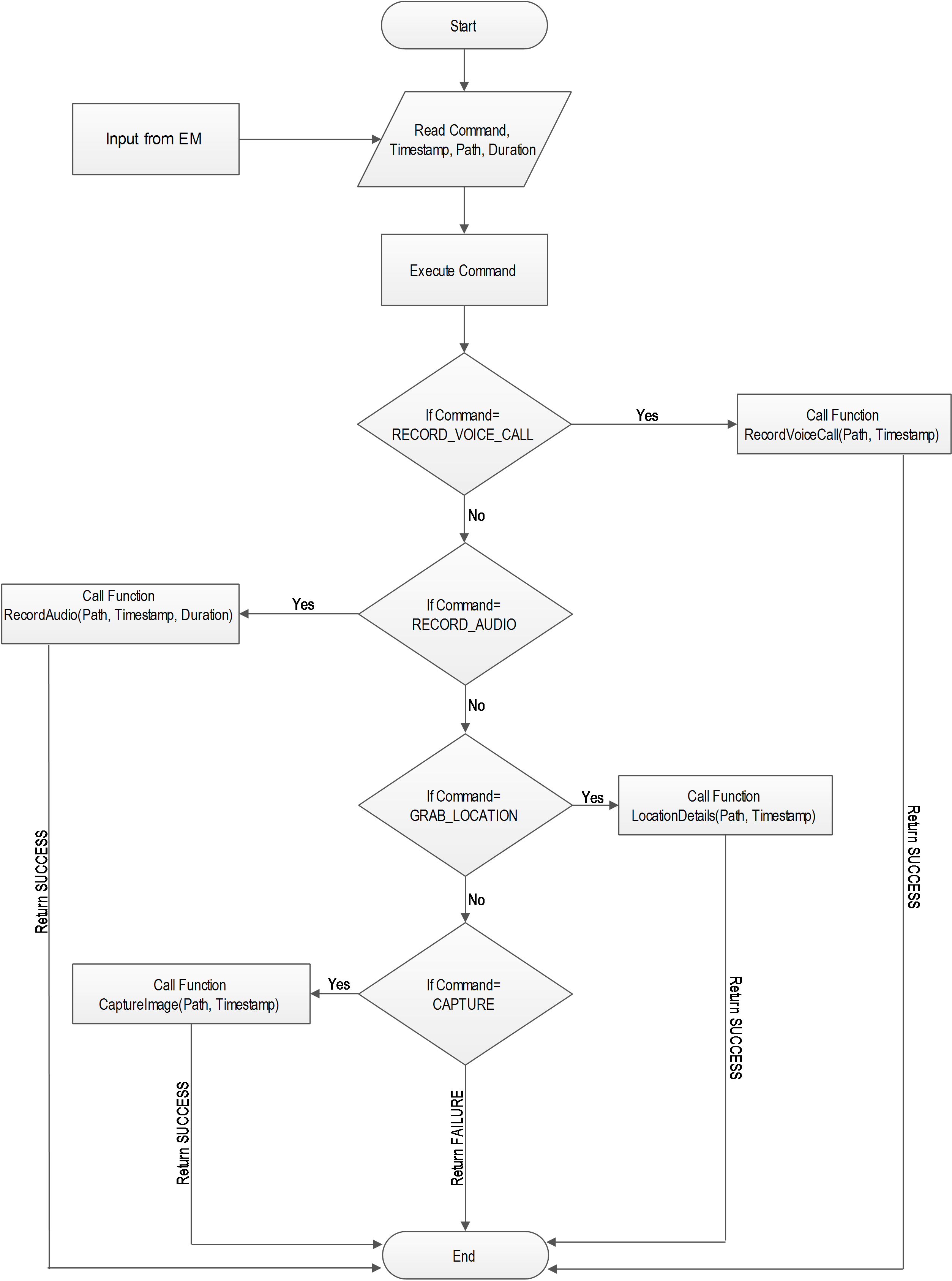}
\caption{Flow Chart for Command Execution Unit}
\label{fig9}
\end{figure}

This unit actually executes the commands. The flow chart in figure \hyperref[fig9]{[Fig. 24]} shows the step-by-step execution of various commands received by the Command Execution Unit.   
\\
The Command Execution Unit starts with receiving a command from the Environment Manager and then it checks which command is to be executed. For instance, if the command received is \textit{RECORD\textunderscore VOICE\textunderscore CALL}, it calls \textit{RecordVoiceCall()} method with parameters \textit{Path}, \textit{Timestamp} and \texttt{Params}. The \texttt{Path} designates the directory in which the output file is to be stored, the \texttt{Timestamp} is the filename uniquely identifying the created file, and \texttt{Params} determines any parameters of that particular command. When the \textit{RecordVoiceCall()} method is called, a new thread gets started which starts a new service \textit{RecordCall} and then registers a broadcast receiver \textit{RVCReceiver}. The parent thread then goes to the waiting state and waits for a notification. The started service \textit{RecordCall} then registers a \textit{Call Broadcast Receiver} and then waits for an incoming or outgoing call to get started. Once a call is initiated on the device, it starts recording the call and when the time value in the \texttt{Params} field expires an \textit{OFF\textunderscore HOOK} event saves the recorded call in the designated folder with a unique name. Then it broadcasts the file to the receiver and stops itself. The receiver on receiving the broadcast notification notifies the parent (\textit{RecordVoiceCall}) thread which then wakes up and unregisters the receiver. It also sends a response back to the calling method. The response is usually \textit{SUCCESS} or \textit{FAILURE} depending upon the status of command executed. This response is redirected to the EM.\\
If the command is \textit{RECORD\textunderscore AUDIO}, the \textit{ExecuteCommand()} method calls the \textit{RecordAudio()} method with three parameters: \texttt{Path}, \texttt{Timestamp}, and \texttt{Params}. The \textit{RecordAudio()} method does not start any new services on the device but simply creates an instance of \textit{Record\textunderscore Audio} class. The \textit{Record\textunderscore Audio} class implements two methods: \textit{startRecord()} and \textit{stopRecord()}. The \textit{startRecord()} method is called by the Command Execution Module (CEM). Once instantiated, it creates an object of \textit{MediaRecorder}, and then sets its various parameters like audio source, and output file using \textit{setAudioSource}, and \textit{setOutputFile} respectively. After setting of parameters is over, it starts the recorder. The \textit{stopRecord()} method is called from child thread spawned by the current thread when the recording time equals the value of the Time in the \texttt{Params} field supplied to the \textit{RecordAudio()} method. The \textit{stopRecord()} method once called, stops, resets, and releases the recorder. The \textit{startRecorder()} method directly returns the result to the calling function as \textit{SUCCESS} or \textit{FAILURE}.\\
If the command is \textit{GRAB\textunderscore GPS\textunderscore LOCATION}, then the \textit{ExecuteCommand()} method calls \textit{GPSLocation()} method with two parameters: \texttt{Path} and \texttt{Timestamp}. When called, the \textit{GPSLocation()} method creates a new service \textit{GrabLocationDetail} and registers a new broadcast receiver \textit{LDReceiver}. The parent thread then goes to the waiting state. The \textit{GrabLocationDetail} service implements a \textit{LocationListener} interface. The \textit{OnCreate()} method associated with this service initializes variables with different values like IsGps to false, lat to 0.0, and lng to 0.0. The \textit{OnStart()} method of this service tries to get the last known location and delivers a broadcast consisting of the coordinates of the current location to the Receiver. The receiver does nothing with these values but waits for another broadcast from the service. The second broadcast is sent from the \textit{OnLocationChanged()} method of the service. When the second broadcast is received by the \textit{Receiver}, it compares its values to the values in the previous broadcast. If there is a change in the values, it notifies the parent thread, which then wakes and writes the given coordinate value to a file. It then unregisters the\textit{ LDReceiver} and returns the result either as \textit{SUCCESS} or as \textit{FAILURE} to the calling method. The calling method, in turn, redirects the result to the Environment Manager (EM) module.   \\
If the command is \textit{CAPTURE\textunderscore IMAGE}, then the \textit{ExecuteCommand()} method calls the \textit{captureImage()} with two parameters. The \textit{captureImage()} method then starts a new service \textit{TakePic} on a new thread and registers a receiver named \textit{CAPReceiver}. The parent thread enters the waiting state. After the \textit{TakePic} service is started, it creates a Camera object. It is this Camera object that opens the camera and takes pictures in the background. After the picture is taken, it is decoded into a bitmap. The bitmap created is saved as a jpg file in the specified directory. After the file is saved, the \textit{TakePic} service sends the broadcast as \textit{SUCCESS} message or otherwise it sends the broadcast as a \textit{FAILURE} message to the \textit{CAPReceiver}. The \textit{CAPReceiver} on receiving the broadcast, forwards the response to the \textit{ExecuteCommand()} method. The \textit{ExecuteCommand()} method redirects the response to the Environment Manager (EM). 

\subsubsection{Data Compressor}
\label{CSdc}

To compress data, we used lossy compression techniques because Android devices today store images and audio in big file sizes and also we do not want the device to swamp the server or use a lot of bandwidth, as in such a case our bot may be discovered by the owner of the device. We used JPEG compression that uses wavelets to compress the images because it is not too complex to run on an Android phone. We use Opus to compress audio because it has a higher compression rate and is most suitable for audio files especially audio files with speech. We didn't compress the text files containing GPS coordinates because the file size is already low. 

\subsubsection{File Upload Unit}
\label{CSfuu}
Now that we are done with the execution of the commands and the data requested is available the bot now tries to upload the data to the server provided the environment to do the same is feasible. This unit implements a single function \textit{UploadFile()} accepting two arguments, the \textit{Path} which is the path of the file to be uploaded, and \textit{URL} which determines the address of the server to upload the data to. It uses the \textit{POST} method to upload the data. The function \textit{UploadFile()} returns a string value to the calling program acknowledging it whether the file has been uploaded or not.

\subsubsection{Sanitizer}
\label{CSsan}
This unit cleanses the bot's data from the storage. Firstly it polls the local database to see if any command has been executed successfully by checking if its status value is \textit{EXECUTED} and if it finds one it deletes its corresponding entry from the table. To delete the data created by this particular command the sanitizer first rewrites garbage data to the associated files and then deletes the same from the storage.\\

\subsection{Experiment setup}
\label{experiment}
The repercussions of the botnet being propagated in the wild are dangerous, therefore we set-up the experiment under a controlled environment because we do not want it to get into the hands of cyber criminals. An application was built for the botmaster through which it disseminates commands and later saves the data. The botmaster also has the ability to balance the number of bots connected to each C\&C server. This application has the capability to archive the whole state of the network continuously to some external device. We used local servers to act as online C\&C servers. We first wrote the infection and inserted it into various famous application's like Truecaller, WhatsApp, Facebook, done via reverse engineering process as in \hyperref[R19]{[19]}. During disassembling an application we made sure that the original working of that application was not hindered. We tested the botnet on about 15 devices with varying features.\\
\end{document}